\documentclass[aps,amsmath,amssymb]{revtex4}
\usepackage{CJK}

\usepackage[latin1]{inputenc}
\usepackage{graphicx}
\usepackage{epstopdf}
\usepackage{bm}
\usepackage{xcolor}
\usepackage{array}
\usepackage{braket}
\usepackage{mathtools}
\usepackage{wasysym}
\renewcommand{\baselinestretch}{1.0}

\begin{document}

\renewcommand{\baselinestretch}{1.0}
\title{Unveiling the Markovian to non-Markovian transition with quantum collision models}

\author{Willames F. Magalh\~{a}es, Carlos O. A. Ribeiro Neto, and Bert\'{u}lio de Lima Bernardo}

\email{bertulio.fisica@gmail.com}

\affiliation{Departamento de F\'{\i}sica, Universidade Federal da Para\'{\i}ba, 58051-900 Jo\~ao Pessoa, PB, Brazil}

\begin{abstract}
  
The concept of non-Markovianity in open quantum systems is traditionally associated with the existence of information backflows from the environment to the system. Meanwhile, the mechanisms through which such backflows emerge are still a subject of debate. In this work, we use collision models to study memory effects in the dynamics of a qubit system in contact with a thermal bath made up of few ancillas, in which system-ancilla and ancilla-ancilla interactions are considered. In the single-ancilla limit case, we show that the system-bath information flow exhibits an interesting mixture of chaotic and regular oscillatory behavior, which depends on the interaction probabilities. In parallel, our results clearly indicate that the information backflows decrease when new ancillas are added to the bath, which sheds light on the nature of the Markovian to non-Markovian transition.

\end{abstract}

\maketitle


\section{Introduction}

Since the past decade, the study of quantum systems interacting with their environment has gained renewed attention in light of the emergence of quantum technologies, such as: quantum computers \cite{ladd,arute,wu}, quantum communication \cite{mishra,simon}, quantum sensing \cite{taylor}, and quantum heat engines \cite{bera,brand,klaers}. In these applications, the main interest lies in finding ways to magnify both the distance over which one reliably transmits a quantum state from one place to another, and the coherence times involved in quantum information processing tasks \cite{nielsen}. In general, the description of the dynamics of open quantum systems is very complicated. However, substantial theoretical progress has been made in understanding and characterizing Markovian  dynamics. This memoryless dynamics approach is considered when the system-environment coupling is weak, and the characteristic times of the system are much longer than those of the environment \cite{breuer,rivas}. 

By contrast, when the open quantum system dynamics is such that the above assumptions do not hold, memory effects become prominent, in the sense that the information that flows from the system to the environment is presumed to flow back to the system after some time \cite{breuer2,vega,cli,cli2}. Investigations of these non-Markovian dynamics are more challenging and have both fundamental and practical importance. From a fundamental viewpoint, this study aims to answer the important question of how quantum correlations between the open system and its environment and correlations within a structured environment influence the system's behavior. From an application viewpoint, it has been reported that the harnessing of non-Markovianity may enhance the efficiency of quantum information processing and communication \cite{byli,white,liu}. 

Many tools can be employed in the study of open quantum systems, which depend on the physical nature of the system-bath interaction properties involved in each case \cite{breuer}. One method that is currently spreading across this research field is the collisional model approach \cite{ciccarello}. In this framework, the bath is considered as a collection of constituents (ancillas), which can individually interact with the system or with another constituent at a time \cite{bernardes}. Here, we use this approach to study the non-Markovian properties of a qubit system that interacts with an environment made up of a small number of thermal ancillas. In the single-ancilla case, we show that memory effects manifest as an interesting mixture of regularity and chaos when varying the interaction probability. It is also shown that these effects are notably suppressed when the number of ancillas increases, which suggests a strong dependence of non-Markovianity on the environment size.

In the following section we present the case in which the system interacts with a single-qubit environment, and examine the dynamics of some information-theoretic properties. Namely, the coherence of the system, and the system-environment entanglement and information flow. In Sec.~III we address the dynamics of the two- and three-qubit environment case, focusing on the change in the non-Markovian behavior as the number of constituents increases, and compare with the Markovian limit in which the environment is made up of an infinitely large number of subunits. Sec.~IV is devoted to our conclusion and discussions.

 \section{Single-qubit environment}\label{s2}

In this section we consider the case in which the open system is a qubit, as is also the case of the environment. This is the limiting case in which the environment is the simplest possible. After presenting the main elements that describe the system-environment dynamics, we then proceed to study the dynamics of the coherence, entanglement and system-environment flow of information. 
   
\subsection{Physical model}
   
\begin{figure}[h!]
 	\includegraphics[scale=0.16]{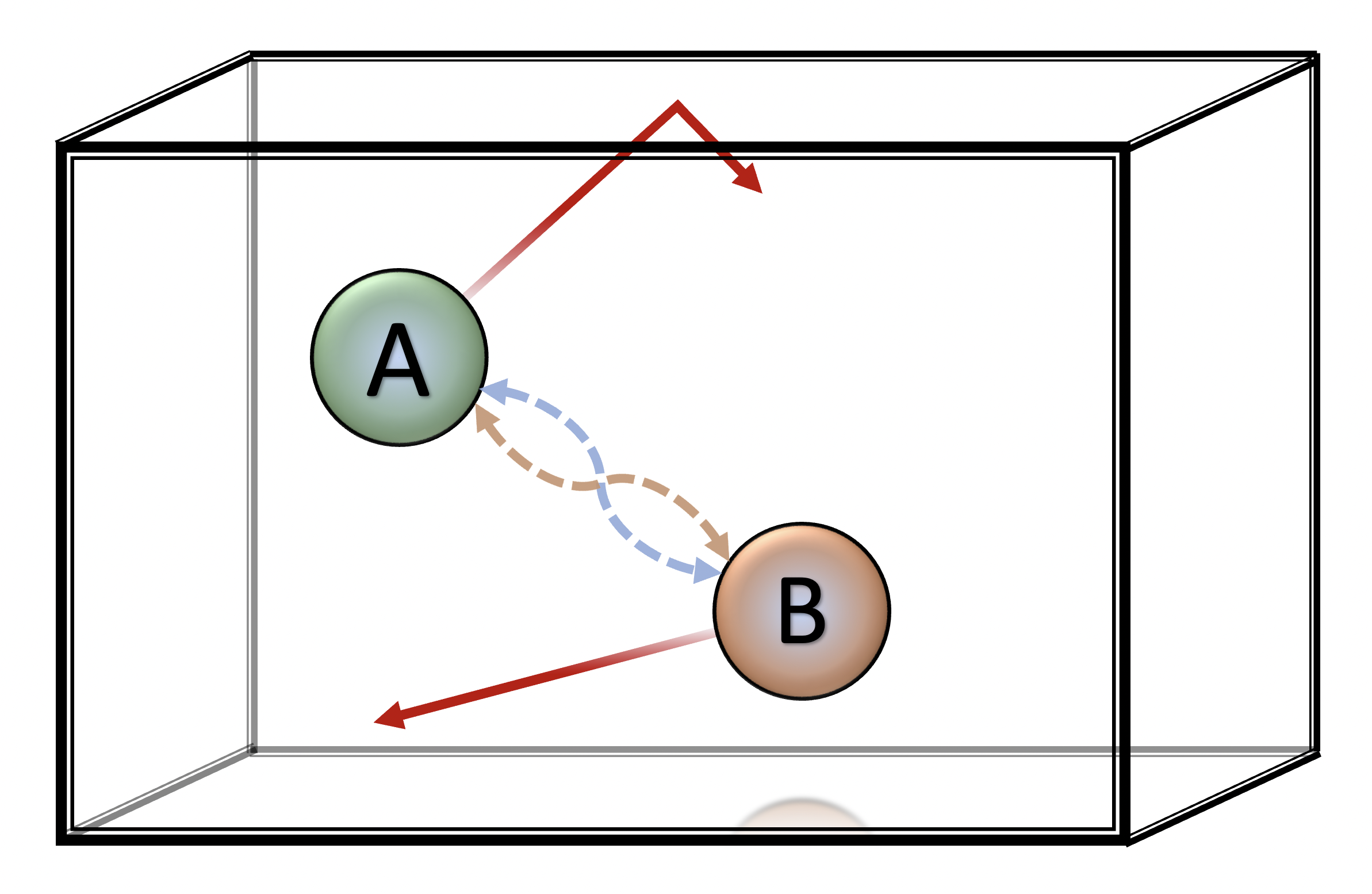}
 	\centering
 	\caption{Pictorial representation of two qubits trapped in a box interacting via a sequence of collisions. Particle $A$ is considered as the main system, and particle $B$ the ancilla.}
 	\label{fig1}
 \end{figure}

Let us consider two qubit particles $A$ and $B$ trapped in a box interacting through a sequence of collisions, as shown in Fig.~\ref{fig1}. Particle $A$ will be treated as our main system and $B$ the environment (ancilla). The ground and excited states of $A$, here denoted by $\ket{g_{A}}$ and $\ket{e_{A}}$, have energies $E_{g}$ and $E_{e}$, respectively. This allows us to write the free Hamiltonian of $A$ as $\hat{H}_{A}= E_{g}\ket{g_{A}}\bra{g_{A}} +E_{e}\ket{e_{A}}\bra{e_{A}}$. Similarly, the ground and excited states of $B$ are represented by $\ket{g_{B}}$ and $\ket{e_{B}}$, with energies $E_{g}$ and $E_{e}$, respectively. With this, we can write $\hat{H}_{B}= E_{g}\ket{g_{B}}\bra{g_{B}} +E_{e}\ket{e_{B}}\bra{e_{B}}$ as the free Hamiltonian of $B$.

To begin with, we assume that $A$ and $B$ are uncorrelated
at the initial time $t=0$,
 \begin{equation}
 \label{initialstate}
 \hat{\rho}_{AB}(0) = \hat{\rho}_{A}(0)\otimes\hat{\rho}_{B}(0), 
 \end{equation}
and that each collision between them is represented by a short unitary interaction $\hat{U}$. As such, after $n$ collisions the state of Eq.~(\ref{initialstate}) is transformed according to the relation 
\begin{equation}
\hat{\rho}_{AB}(0)\rightarrow
\hat{\rho}_{AB}(n)=\hat{U}^{n}\hat{\rho}_{A}(0)\otimes\hat{\rho}_{B}(0)\hat{U}^{\dagger n}. 
\end{equation}
If we are interested in obtaining information about $A$ and $B$, this is where their reduced density matrices takes place,
\begin{equation}
\hat{\rho}_{A}(n)=Tr_{B}[\hat{U}^{n}\hat{\rho}_{A}(0)\otimes\hat{\rho}_{B}(0)\hat{U}^{\dagger n}] 
\end{equation}
and
\begin{equation}
\hat{\rho}_{B}(n)=Tr_{A}[\hat{U}^{n}\hat{\rho}_{A}(0)\otimes\hat{\rho}_{B}(0)\hat{U}^{\dagger n}],
\end{equation}
where $Tr_{A}$ and $Tr_{B}$ represent the trace over the states of $A$ and $B$, respectively.

As an example, let us assume the initial state of $A$ as the pure state
\begin{equation}
\label{stateA0}
\ket{\psi(0)} = a\ket{g_{A}}+b\ket{e_{A}},
\end{equation}
with $|a|^2+|b|^2 = 1$. This provides
\begin{equation}
\label{idensmatA}
\begin{split}
\hat{\rho}_{A}(0) & = \ket{\psi(0)}\bra{\psi(0)}\\
&=|a|^{2}\ket{g_{A}}\bra{g_{A}} + ab^{*}\ket{g_{A}}\bra{e_{A}} \\
&+ a^ {*}b\ket{e_{A}}\bra{g_{A}}+|b|^{2}\ket{e_{A}}\bra{e_{A}}.
\end{split}
\end{equation}
The initial state of $B$ is assumed to be a thermal state with inverse temperature $\beta$,
\begin{equation}
\label{idensmatB}
\hat{\rho}_{B}(0) = w_{g}\ket{g_{B}}\bra{g_{B}} + w_{e}\ket{e_{B}}\bra{e_{B}},
\end{equation}
such that the probabilistic weights obey the relation $w_{e} = w_{g} e^{-\beta(E_{e} - E_{g})}$, with $w_{g}+w_{e} = 1$. Having defined the density matrices in Eqs.~(\ref{idensmatA}) and (\ref{idensmatB}), from Eq.~(\ref{initialstate}) we can write the initial density operator of the composite system in the basis $\{ \ket{g_{A},g_{B}},\ket{g_{A},e_{B}},\ket{e_{A},g_{B}}, \ket{e_{A},e_{B}} \}$ as given by
\begin{equation}\label{d99}
\hat{\rho}_{AB}(0)=\begin{pmatrix}
|a|^{2}w_{g}&0&ab^{*}w_{g}&0\\
0&|a|^{2}w_{e}&0&ab^{*}w_{e}\\
a^{*}bw_{g}&0&|b|^{2}w_{g}&0\\
0&a^{*}bw_{e}&0&|b|^{2}w_{e}\\
\end{pmatrix},
\end{equation}
where the asterisk denotes complex conjugation.

Next, we describe the unitary interactions that result from the collisions.
A natural choice in this case is to use a unitary transformation that makes the system to evolve according to the generalized amplitude-damping channel (GADC), which models the dynamics of a qubit in contact with a thermal reservoir at nonzero temperature \cite{khatri}. In fact, the GADC has been used to model
a spin-1/2 particle coupled to an interacting spin chain \cite{bose}, to characterize noise in
superconducting-circuit-based quantum computing \cite{goold,chirolli}, thermal noise in linear optical systems \cite{zou}, and the thermodynamics of a two-level system strongly coupled to a thermal environment \cite{bernardo}. To implement this idea we consider the following system-environment dynamics:

\begin{equation}\label{w1}
\ket{g_{A},g_{B}}\longrightarrow\ket{g_{A},g_{B}},
\end{equation}
\begin{equation}\label{w2}
\ket{g_{A},e_{B}}\longrightarrow\sqrt{1-p}\ket{g_{A},e_{B}} -\sqrt{p}\ket{e_{A},g_{B}},
\end{equation}
\begin{equation}\label{w3}
\ket{e_{A},g_{B}}\longrightarrow\sqrt{1-p}\ket{e_{A},g_{B}} +\sqrt{p}\ket{g_{A},e_{B}},
\end{equation}
\begin{equation}\label{w4}
\ket{e_{A},e_{B}}\longrightarrow\ket{e_{A},e_{B}}.
\end{equation}

From this set of relations we can make some observations. {\it i}) Eq.~(\ref{w1}) means that if both $A$ and $B$ are in the ground state, no transition occurs. {\it ii}) Eq.~(\ref{w2}) tells us that if $A$ is in the ground state and $B$ in the excited state,
after one collision there is a probability $p$ that $A$ becomes excited and $B$ decays to the ground state. {\it iii}) Eq.~(\ref{w3}) says that if $A$ and $B$ are respectively in the excited and ground states, after one collision there is a probability $p$ of $A$ decaying to the ground state and $B$ making a transition to the excited state. Finally, {\it iv}) Eq.~(\ref{w4}) indicates that if $B$ is excited, the excited state of $A$ has a lifetime which is much longer than the characteristic time between successive collisions. This means that we are assuming that the state $\ket{e_{A}}$ becomes metastable when the ancilla is in the excited state \cite{maci,boite,valenti}.

As a result, from Eqs.~(\ref{w1}) to~(\ref{w4}), we can construct the unitary matrix $\hat{U}$ that describes the collisional interactions as
\begin{equation}\label{canal1}
\hat{U}=\begin{pmatrix}
1&0&0&0\\
0&\sqrt{1-p}&\sqrt{p}&0\\
0&-\sqrt{p}&\sqrt{1-p}&0\\
0&0&0&1\\
\end{pmatrix},
\end{equation}
with $p \in [0,1]$.
Having presented this first model, in what follows we study some information-theoretic properties of the dynamics of $A$ and $B$, and the flow of information between them.

\subsection{Coherence Dynamics}

Important information about the system-environment dynamics can be revealed if one studies the coherence dynamics of the system \cite{bernardo2,khan}. In this regard, we study how the coherence of $A$ changes as a function of the multiple collisions with $B$. Here, we quantify coherence with the $l_{1}$ norm of coherence \cite{baum}, 
\begin{equation}\label{norma}
C_{A}(n)=\sum_{i\neq j}|\rho^{A}_{ij}(n)|,
\end{equation}
where $\rho^{A}_{ij}(n)$ represent the off-diagonal elements of the reduced density matrix of $A$ after the $n$th collision. We assume the initial state of the system as $\ket{\psi_{A}(0)} = 1/ \sqrt{2} (\ket{g_{A}}+\ket{e_{A}})$, which is maximally coherent, $\mathcal{C}_{A}(0) = 1$. For the ancilla we consider the initial thermal state of Eq.~(\ref{idensmatB}) with $w_{g} = 0.8$ and $w_{e} = 0.2$, which corresponds to $\beta = \ln(4)/(E_{e} - E_{g})$.

\begin{figure}[h!]
    \includegraphics[scale=0.46]{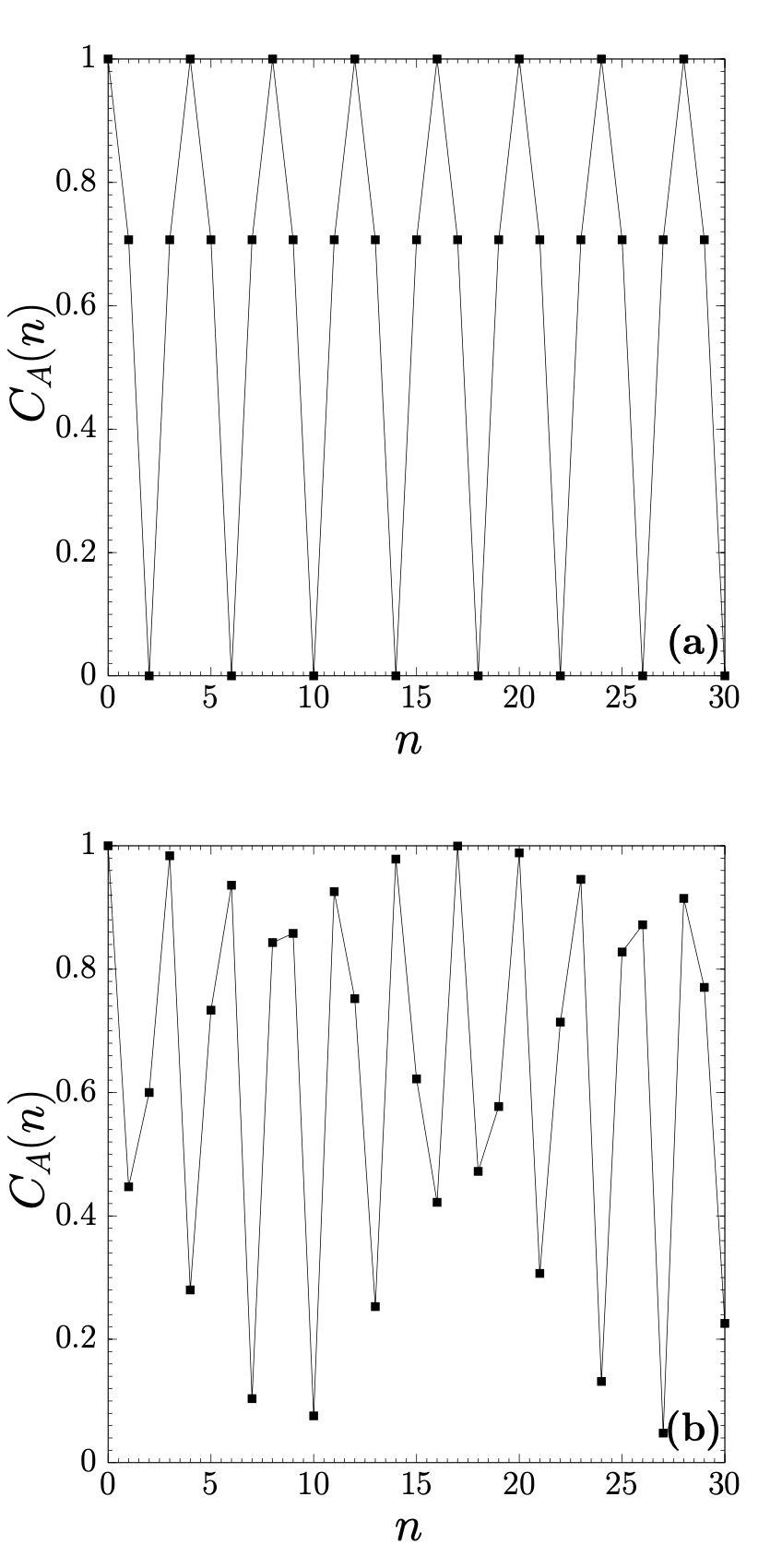}
	\centering
	\caption{Coherence of the system $A$ as a function of the number of collisions with the ancilla $B$. $A$ is assumed to be initially in the maximally coherent state $\ket{\psi_{A}(0)} = 1/ \sqrt{2} (\ket{g_{A}}+\ket{e_{A}})$, and $B$ in the thermal state $\rho_{B}(0) =  0.8\ket{g_{B}}\bra{g_{B}} + 0.2\ket{e_{B}}\bra{e_{B}}$. After each collision, $A$ and $B$ are transformed according to the unitary operation of Eq.~(\ref{canal1}), where we defined the interaction probabilities: (a) $p=0.5$ and (b) $p=0.8$. We observe that the coherence dynamics may be either regular or chaotic.}
	  \label{fig2}
\end{figure}

In Fig.~\ref{fig2}(a) we plot the results of ${C}_{A}(n)$ versus $n$ as a {\it time series} for $p=0.5$. For clarity, we used line segments to connect the discrete points $[\mathcal{C}_{A}(n),n]$, which are the only meaningful results. As can be seen, the coherence oscillates alternating from 0 to 1, assuming only three values: $C_{A}=0,1/\sqrt{2},1$. Namely, the system experiences decoherence and recoherence (recovery of coherence) indefinitely, and never attains a steady state. This type of regular oscillation, in which a given property repeats every {\it four} iterations is named {\it period-4 cycle} \cite{strogatz}. In Fig.~\ref{fig2}(b) we show the results of ${C}_{A}(n)$ versus $n$ for $p=0.8$. In this case, we also have an alternation between decoherence and recoherence, but this behavior never settles down to a periodic orbit as in the $p=0.5$ case. Instead, we observe an aperiodic (chaotic) long-term behavior. In the next subsection we will see how this process of coherence exchange between $A$ and $B$ relates with the entanglement generation between them. For now, we note that the value of $p$ determines the behavior of the coherence dynamics. 

\begin{figure}[h!]
	\includegraphics[scale=0.67]{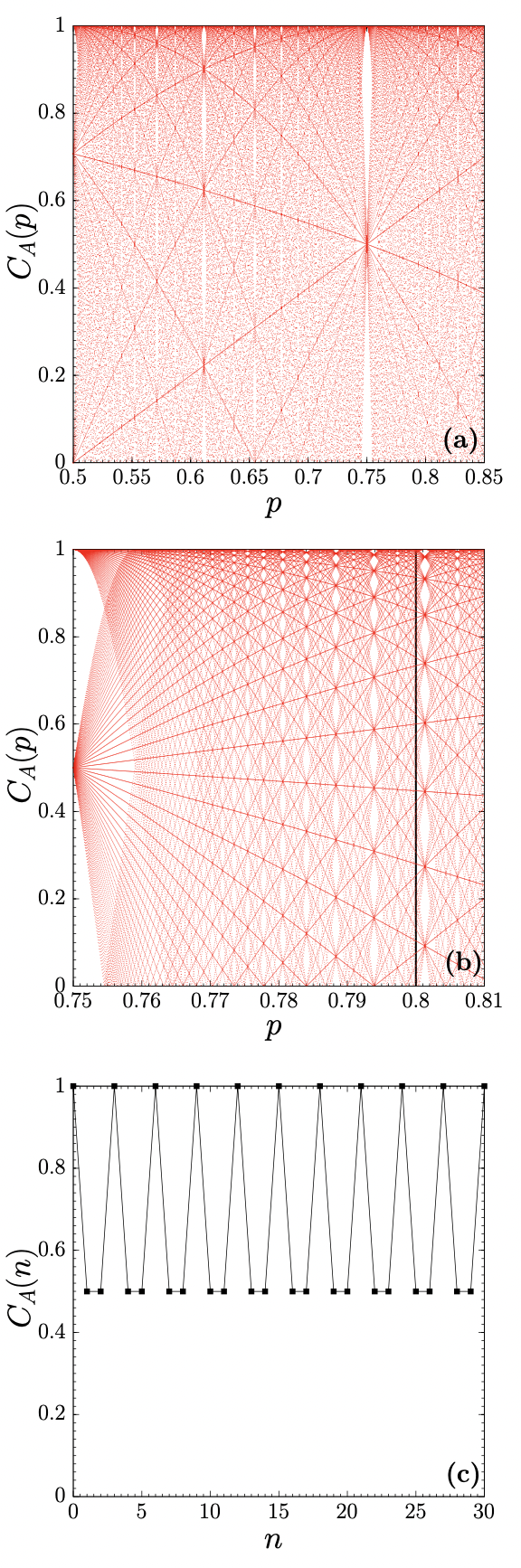}
	\centering
	\caption{(a) Orbit diagram of the coherence of the system $A$ as a function of the interaction probability $p$ with the ancilla $B$. The diagrams were constructed for $100$ collisions, with the initial states of $A$ and $B$ given by $\ket{\psi_{A}(0)} = 1/ \sqrt{2} (\ket{g_{A}}+\ket{e_{A}})$ and $\rho_{B}(0) =  0.8\ket{g_{B}}\bra{g_{B}} + 0.2\ket{e_{B}}\bra{e_{B}}$, respectively. A blow-up of (a) is shown in (b) with a black vertical line located at $p=0.8$. (c) Coherence of $A$ as a function of the number of collisions for $p=0.75$. We observe that the probability interaction $p$ determines whether the coherence dynamics is regular or chaotic.}
	\label{fig3}
\end{figure}

A natural question is whether the coherence of the system presents other types of dynamics as the value of $p$ varies. To answer this question, in Fig.~\ref{fig3} we display {\it orbit diagrams} of the coherence dynamics, which show the long-term behavior for a continuum of values of $p$ at once. Fig.~\ref{fig3}(a) shows the results for the interval $0.5 \leq p \leq 0.85$. At $p = 0.5$ we see the period-4 cycle pointed out above with the three possible values. As $p$ increases, the periodicity rapidly disappears, giving place to a chaotic behavior in which coherence can assume an infinite set of points. Nevertheless, a number of other periodic regions show up as $p$ increases, so that the orbit diagram reveals an interesting mixture of regularity and chaos, with {\it periodic apertures} separated by chaotic intervals. For example, a period-5 cycle can be seen at $p \approx 0.62$, and a very prominent period-3 cycle at $p=0.75$. Fig.~\ref{fig3}(b) shows a zoom of the diagram in the region $0.75 \leq p \leq 0.85$, in which we can observe an interesting structure of points containing a group of high-order period cycles. We added a vertical line at $p=0.8$ to indicate the disordered behavior exhibited in Fig.~\ref{fig2}(b). Fig.~\ref{fig3}(c) displays the time series of the period-3 cycle of ${C}_{A}(n)$ for $p=0.75$, in which the system never becomes totally incoherent.

\subsection{Entanglement Dynamics}

Having described the coherence dynamics of $A$, we now proceed to examine the coherence of $B$ along with the system-environment entanglement dynamics which is formed as a result of the multiple collisions. Since the collisions are described by qubit-qubit interactions, we can use the concept of negativity to quantify the entanglement. The negativity is an entanglement monotone given by \cite{vidal,horodecki}
\begin{equation}\label{jj99}
\mathcal{N}[\hat{\rho}_{AB}(n)]=\frac{\|\hat{\rho}^{T_{A}}_{AB}(n)\|_{1}-1}{2},
\end{equation}
where $\hat{\rho}_{AB}(n)= \hat{U}^{n}\hat{\rho}_{AB}(0)\hat{U}^{\dagger n}$ denotes the density matrix of the composite system after the $n$th collision, and $\hat{\rho}_{AB}^{T_{A}}(n)$ the partial transpose of $\hat{\rho}_{AB}(n)$ with respect to $A$. The trace norm of an operator $\hat{O}$ is defined as $\| \hat{O} \| = $ tr$\{ \sqrt{\hat{O} \hat{O}^{\dagger}} \}$. The negativity can also be written as the sum of the absolute values of the negative eigenvalues of $\hat{\rho}_{AB}^{T_{A}}(n)$, which is null for separable states.

\begin{figure*}
\includegraphics[scale=0.28]{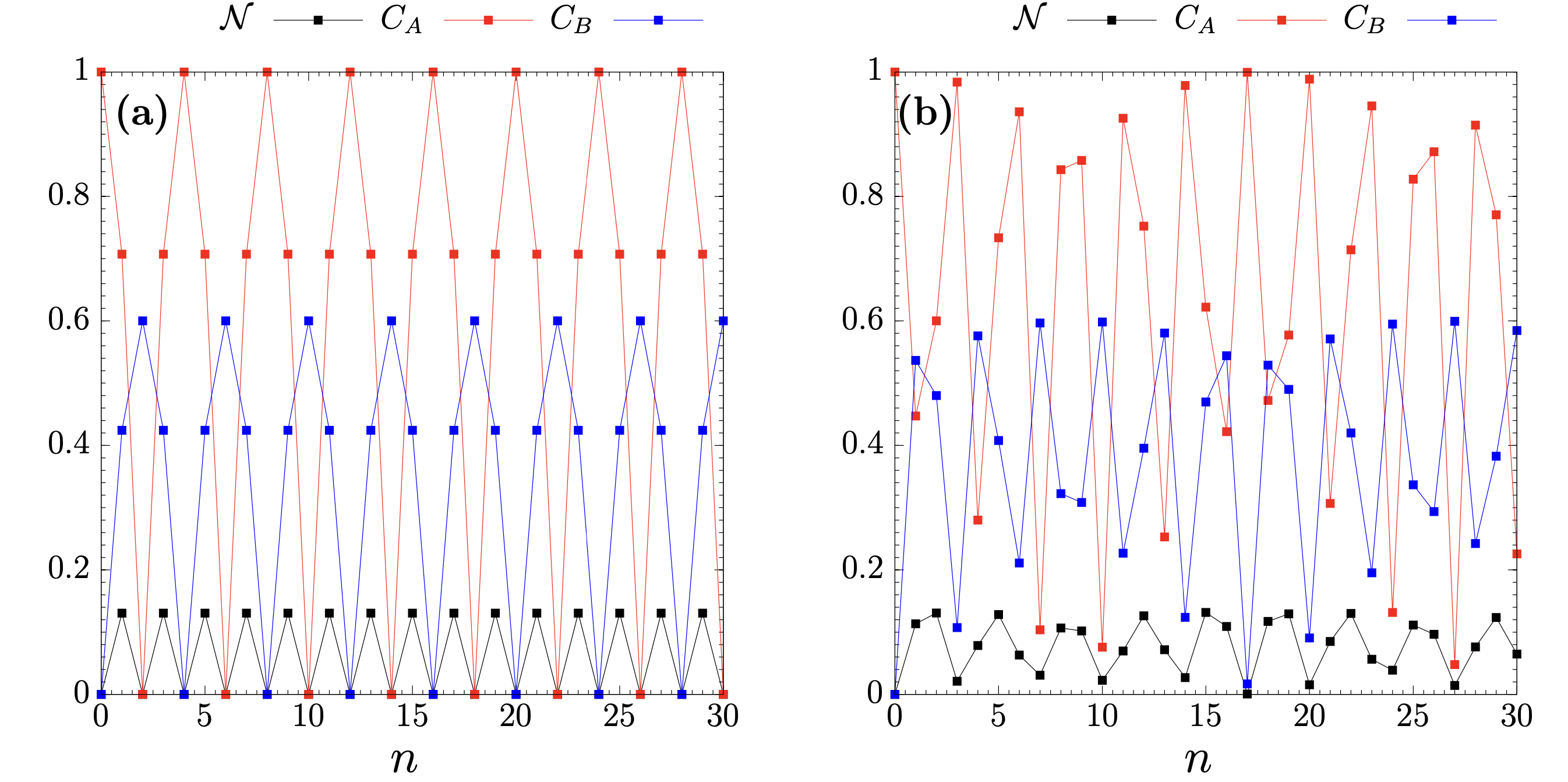}
\centering
\caption{Behavior of the entanglement negativity $\mathcal{N}$ (black curve), and $l_{1}$ norm of coherence $C_{A}$ (red curve) and $C_{A}$ (blue curve) as a function of the number of collisions $n$ between $A$ and $B$, where we used (a) $p=0.5$ and (b) $p=0.8$. We observe that, similar to the coherence dynamics, the entanglement dynamics may be either regular or chaotic.}
\label{fig4}
\end{figure*}

In Fig.~\ref{fig4}(a) we observe that for $p=0.5$ the entanglement between $A$ and $B$ oscillates between the values $0$ and approximately $0.13$ in a period-2 cycle. This means that the composite system alternates between entangled and separable states. To further understand this dynamics, we also show the behavior of the coherence dynamics of $B$ which, following the example of $A$, also exhibits a period-4 cycle. In the absence of entanglement, the coherence of $A$ is either maximal (with $B$ in a incoherent state), or null (with $B$ in a partially coherent state, ${C}_{B} = 0.6$). On the other hand, whenever $A$ and $B$ are partially entangled, $\mathcal{N}(\hat{\rho}_{AB}) \approx 0.13$, they manifest partial coherence, ${C}_{A} \approx 0.71$ and ${C}_{B} \approx 0.42$. Fig.~\ref{fig4}(b) shows the results of $\mathcal{N}(\hat{\rho}_{AB})$, ${C}_{A}$ and ${C}_{B}$ for $p=0.8$. Unlike the $p=0.5$ case, the three quantities present chaotic oscillatory behavior. As expected, we see that nonzero entanglement implies partial coherence of both $A$ and $B$.

\subsection{System-environment information flow}

One of the main problems in describing non-Markovianity in the quantum realm lies in defining how memory effects manifest at this level. As pointed out in the introduction, what we understand by quantum non-Markovianity is the dynamical property in which the information that flowed from the system to the environment, after some time, flows totally or partially back to the system \cite{breuer2,vega,cli,cli2}. One of the first proposals to characterize information flow between system and environment to quantify quantum non-Markovianity used the concept of trace distance $\mathcal{D}(\hat{\rho}_{1},\hat{\rho}_{2})$ \cite{breuer3}, which is a measure of the distinguishability between two arbitrary quantum states, $\hat{\rho}_{1}$ and $\hat{\rho}_{2}$ \cite{nielsen}. The idea is that, in a non-Markovian dynamics of a system $A$, there is a pair of initial states, $\hat{\rho}_{A}^{(1)}(0)$ and $\hat{\rho}_{A}^{(2)}(0)$, such that for a certain time interval $\Delta t = t_{f} - t_{i}>0$ the distinguishability increases as a consequence of the backflow of information from the environment,
\begin{eqnarray}
\label{td}
\Delta \mathcal{D}[\hat{\rho}_{A}^{(1)}(t),\hat{\rho}_{A}^{(2)}(t)]&=&\frac{1}{2} \Delta \|\hat{\rho}_{A}^{(1)}(t)-\hat{\rho}_{A}^{(2)}(t)\| \nonumber \\
&=& \frac{1}{2} \Delta \sum_{i}|\lambda_{i}(t)| >0,
\end{eqnarray}
where for a general function of time $f(t)$ we have $\Delta f(t) = f(t_{f}) - f(t_{i})$, $\|\hat{A}\| = \sqrt{\hat{A}^\dag \hat{A}}$ represents the absolute value of an operator $\hat{A}$, and $\lambda_{i}(t)$ the eigenvalues of $\hat{\rho}_{A}^{(1)}(t)-\hat{\rho}_{A}^{(2)}(t)$. The second equality of Ineq. (\ref{td}) holds because $\hat{\rho}_{A}^{(1)}(t)-\hat{\rho}_{A}^{(2)}(t)$ is a time-dependent Hermitian operator \cite{nielsen}.

\begin{figure}[h!]
	\includegraphics[scale=0.47]{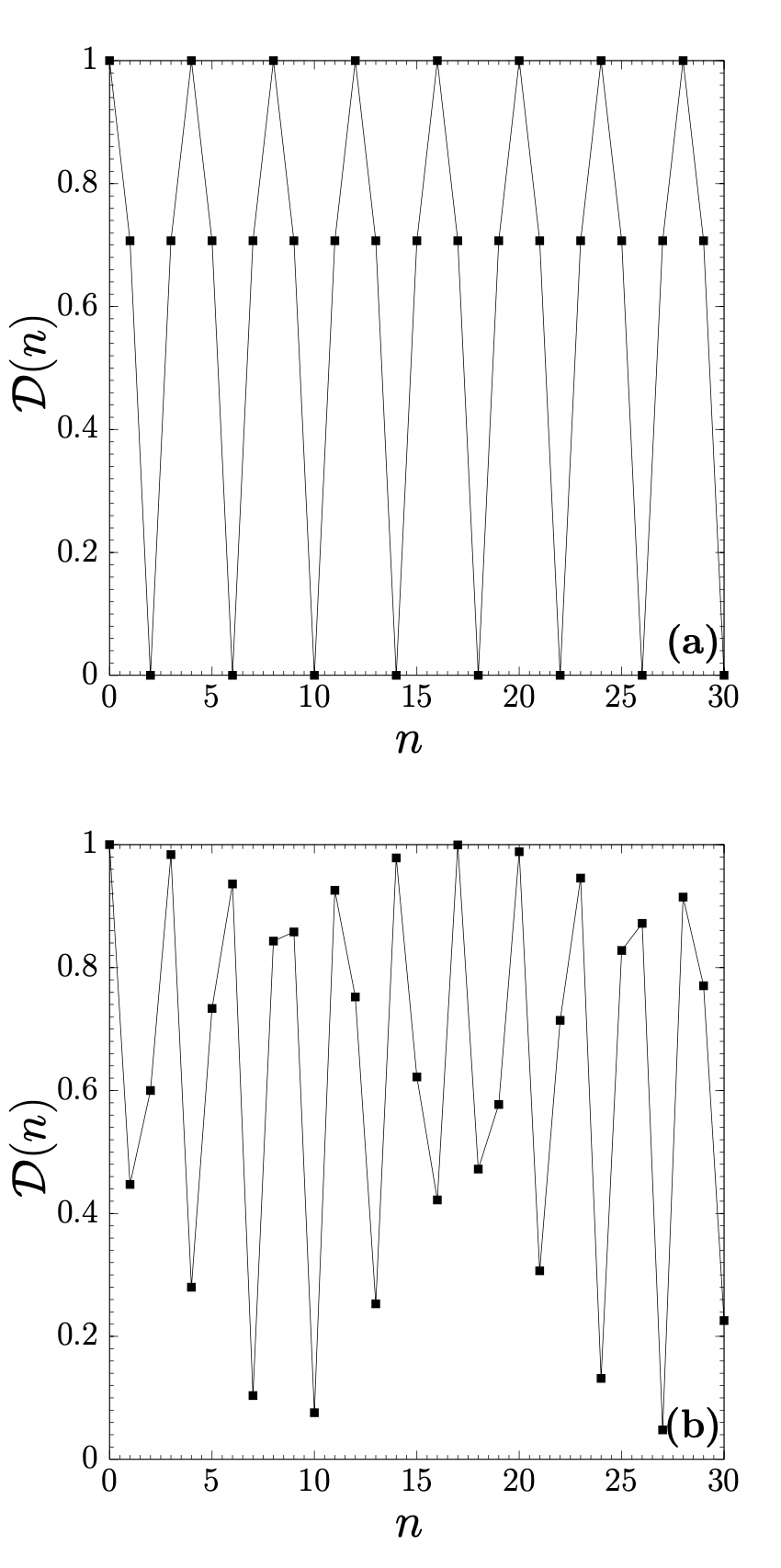}
	\centering
	\caption{Trace distance of the initially orthogonal states of $A$ as a function of the number of collisions. For the initial state of the single ancilla $B$ we used Eq.~(\ref{idensmatB}) with $w_{g}=0.8$ and $w_{e}=0.2$. We defined (a) $p=0.5$ and (b) $p=0.8$. In both cases, the persistent oscillatory behavior indicates strong non-Markovianity. }
	\label{fig5}
\end{figure}

Next, we use the above criterion to investigate the presence of non-Markovianity in the dynamics of $A$ and $B$. We assume two initial distinguishable (orthogonal) states of the system, numerically calculate the evolution of these states as a result of the multiple collisions with the ancilla, and observe if a given collision could cause an increase in the trace distance of the evolved states. If so, we may conclude that the dynamics is non-Markovian. We consider the two initial states  
\begin{equation}\label{rho1}
\ket{\psi_{A}^{(1)}(0)} = \frac{1}{\sqrt{2}}(\ket{g_{A}}+\ket{e_{A}})
\end{equation}
and
\begin{equation}\label{rho2}
\ket{\psi_{A}^{(2)}(0)} = \frac{1}{\sqrt{2}}(\ket{g_{A}}-\ket{e_{A}})
\end{equation}
to be used to verify the criterion of Ineq.~(\ref{td}). To this end, we calculate the value of $\mathcal{D}(n)$, which is a simplified notation for the trace distance between the evolved states after the $n$th collision. The results are shown in Fig.~\ref{fig5}. 

We observe that $\mathcal{D}(n)$ presents a persistent oscillatory behavior with infinitely many evolution steps satisfying Ineq.~(\ref{td}). These evolution steps with increasing distinguishability characterize the presence of memory effects in the quantum dynamics. In Fig.~\ref{fig5}(a) we see the results for $p=0.5$, in which the trace distance presents a period-4 cycle, assuming only three values: $\mathcal{D}(n)=0,1/\sqrt{2},1$. As found in the coherence and entanglement analyses, this result also confirms the regularity of the dynamical properties of the system when $p=0.5$. In Fig.~\ref{fig5}(b) we show the results for $p=0.8$, where we also observe infinitely many evolution steps satisfying Ineq.~(\ref{td}). 
However, the trace distance presents a chaotic oscillatory behavior, which corroborates our previous findings about the irregularity of the quantum dynamics revealed by the study of the coherence and entanglement. The strong signature of non-Markovianity verified in this two-qubit collision model is justified by the fact that, since $A$ and $B$ have the same size and dimensionality, it is natural that any information that flows from one to the other eventually returns. From now on, we will focus on the influence of the environment size on the manifestation of memory effects.

\section{Two- and three-qubit environment}
\begin{figure}[h!]
	\includegraphics[scale=0.16]{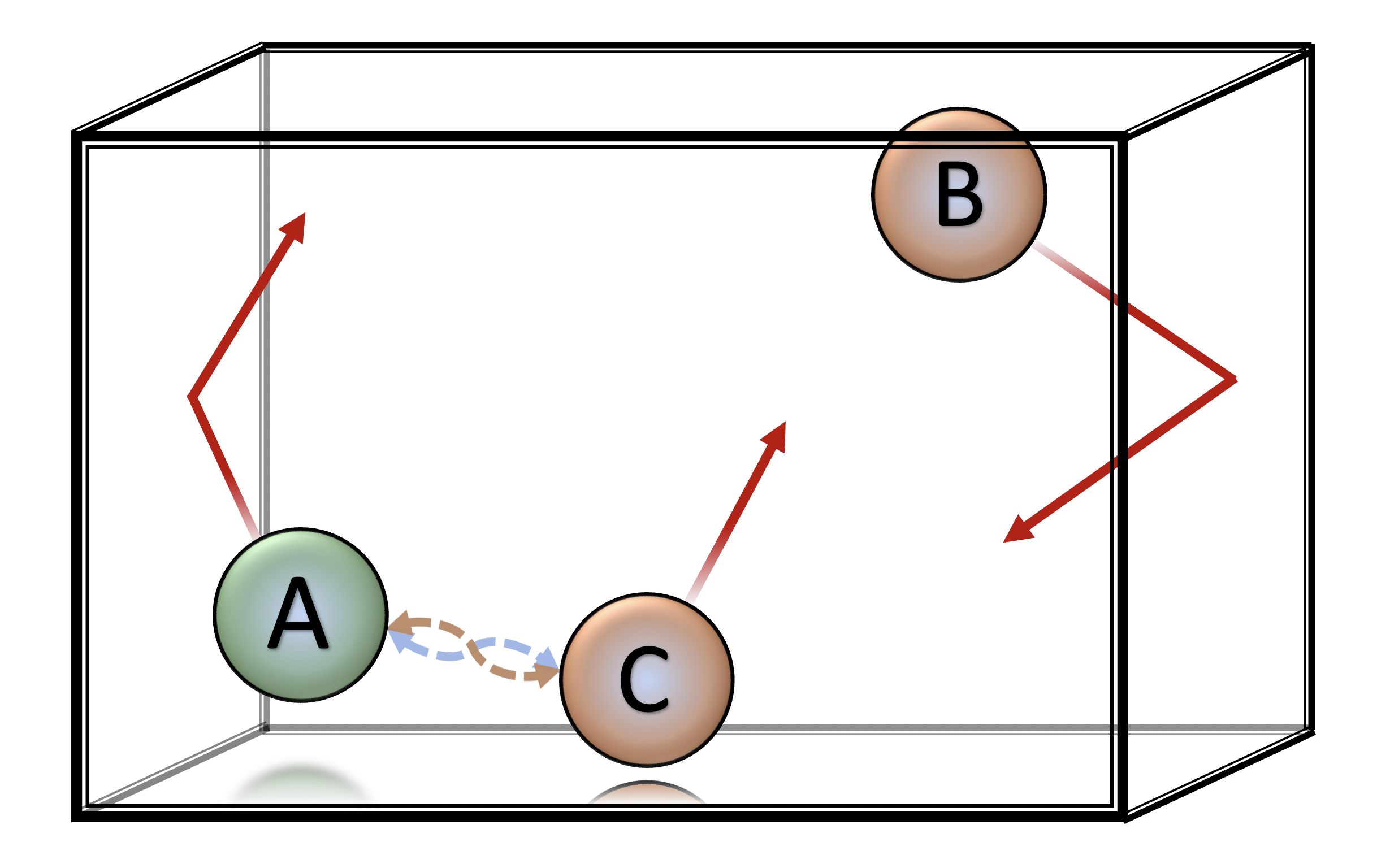}
	\centering
	\caption{Pictorial representation of three qubits trapped in a box interacting via a sequence of collisions. Particle $A$ is considered as the main system, and particles $B$ and $C$ the environment.}
	\label{fig6}
\end{figure}

Let us now study how the non-Markovian effects change when the environment size increases. We begin with by adding a qubit $C$ to the environment, whose initial state is identical to that of $B$, i.e., $\hat{\rho}_{B}(0) = w_{g}\ket{g_{B}}\bra{g_{B}} + w_{e}\ket{e_{B}}\bra{e_{B}}$ and $\hat{\rho}_{C}(0) = w_{g}\ket{g_{C}}\bra{g_{C}} + w_{e}\ket{e_{C}}\bra{e_{C}}$. We suppose that $A$, $B$ and $C$ start out in a separable state with $\hat{\rho}_{A}(0)$ given as in Eq.~(\ref{stateA0}), that they can only interact in pairs in a sequence of many collisions, and that all qubit-qubit interactions are described by the relations shown in Eqs.~(\ref{w1}) to~(\ref{w4}). In Fig.~\ref{fig6} we show a pictorial representation of this arrangement, in which collisions take place randomly. In what follows, we depict how the three-particle state evolves when a given collision occurs. Since quantum correlations involving the three particles arise upon evolution, the qubit-qubit collisions must be described in the energy Hilbert space of $A$, $B$ and $C$. This space can be spanned by a set of energy states as: $\{ \ket{g_{A},g_{B},g_{C}}, \ket{g_{A},g_{B},e_{C}}, \ket{g_{A},e_{B},g_{C}}, \ket{g_{A},e_{B},e_{C}},$
$\ket{e_{A},g_{B},g_{C}}, \ket{e_{A},g_{B},e_{C}}, \ket{e_{A},e_{B},g_{C}}, \ket{e_{A},e_{B},e_{C}}\}$. In this scenario, if we are interested for example in the evolution of the three particles when $A$ collides with $B$, we must extend the relations~(\ref{w1}) to~(\ref{w4}) to the form 

\begin{equation}\label{w5}
\ket{g_{A},g_{B},g_{C}} \longrightarrow \ket{g_{A},g_{B},g_{C}},
\end{equation}
\begin{equation}\label{w6}
\ket{g_{A},g_{B},e_{C}} \longrightarrow \ket{g_{A},g_{B},e_{C}},
\end{equation}
\begin{equation}\label{w7}
\ket{g_{A},e_{B},g_{C}} \longrightarrow\sqrt{1-p} \ket{g_{A},e_{B},g_{C}} -\sqrt{p} \ket{e_{A},g_{B},g_{C}},
\end{equation}
\begin{equation}\label{w8}
\ket{g_{A},e_{B},e_{C}} \longrightarrow\sqrt{1-p} \ket{g_{A},e_{B},e_{C}} -\sqrt{p} \ket{e_{A},g_{B},e_{C}},
\end{equation}
\begin{equation}\label{w9}
\ket{e_{A},g_{B},g_{C}} \longrightarrow\sqrt{1-p} \ket{e_{A},g_{B},g_{C}} +\sqrt{p} \ket{g_{A},e_{B},g_{C}},
\end{equation}
\begin{equation}\label{w10}
\ket{e_{A},g_{B},e_{C}} \longrightarrow\sqrt{1-p} \ket{e_{A},g_{B},e_{C}} +\sqrt{p} \ket{g_{A},e_{B},e_{C}},
\end{equation}
\begin{equation}\label{w11}
\ket{e_{A},e_{B},g_{C}} \longrightarrow \ket{e_{A},e_{B},g_{C}},
\end{equation}
\begin{equation}\label{w12}
\ket{e_{A},e_{B},e_{C}} \longrightarrow \ket{e_{A},e_{B},e_{C}}.
\end{equation}
As can be seen, the energy states of particle $C$, $\ket{g_{C}}$ and $\ket{e_{C}}$, have no influence on this dynamics. With the above relations we can write the unitary matrix that describes a collision between $A$ and $B$:

\begin{equation}\label{canal2}
\hat{U}_{AB}=\begin{pmatrix}
1&0&0&0&0&0&0&0\\ 0&1&0&0&0&0&0&0\\
0&0&\sqrt{1-p}&0&\sqrt{p}&0&0&0\\
0&0&0&\sqrt{1-p}&0&\sqrt{p}&0&0\\
0&0&-\sqrt{p}&0&\sqrt{1-p}&0&0&0\\
0&0&0&-\sqrt{p}&0&\sqrt{1-p}&0&0\\
0&0&0&0&0&0&1&0\\
0&0&0&0&0&0&0&1\\

\end{pmatrix}.
\end{equation}

In a similar fashion, we can describe the dynamics of a collision between $A$ and $C$. This is given by the relations

\begin{equation}\label{w13}
\ket{g_{A},g_{B},g_{C}} \longrightarrow \ket{g_{A},g_{B},g_{C}},
\end{equation}
\begin{equation}\label{w14}
\ket{g_{A},g_{B},e_{C}} \longrightarrow\sqrt{1-p} \ket{g_{A},g_{B},e_{C}} -\sqrt{p} \ket{e_{A},g_{B},g_{C}},
\end{equation}
\begin{equation}\label{w15}
\ket{g_{A},e_{B},g_{C}} \longrightarrow \ket{g_{A},e_{B},g_{C}},
\end{equation}
\begin{equation}\label{w16}
\ket{g_{A},e_{B},e_{C}} \longrightarrow\sqrt{1-p} \ket{g_{A},e_{B},e_{C}} -\sqrt{p} \ket{e_{A},e_{B},g_{C}},
\end{equation}
\begin{equation}\label{w17}
\ket{e_{A},g_{B},g_{C}} \longrightarrow\sqrt{1-p} \ket{e_{A},g_{B},g_{C}} +\sqrt{p} \ket{g_{A},g_{B},e_{C}},
\end{equation}
\begin{equation}\label{w18}
\ket{e_{A},g_{B},e_{C}} \longrightarrow \ket{e_{A},g_{B},e_{C}} ,
\end{equation}
\begin{equation}\label{w19}
\ket{e_{A},e_{B},g_{C}} \longrightarrow\sqrt{1-p} \ket{e_{A},e_{B},g_{C}} + \sqrt{p}\ket{g_{A},e_{B},e_{C}},
\end{equation}
\begin{equation}\label{w20}
\ket{e_{A},e_{B},e_{C}} \longrightarrow \ket{e_{A},e_{B},e_{C}}.
\end{equation}
These relations provide the following unitary evolution matrix:

\begin{equation}\label{canal3}
\hat{U}_{AC}=\begin{pmatrix}
1&0&0&0&0&0&0&0\\ 0&\sqrt{1-p}&0&0&\sqrt{p}&0&0&0\\
0&0&1&0&0&0&0&0\\
0&0&0&\sqrt{1-p}&0&0&\sqrt{p}&0\\
0&-\sqrt{p}&0&0&\sqrt{1-p}&0&0&0\\
0&0&0&0&0&1&0&0\\
0&0&0&-\sqrt{p}&0&0&\sqrt{1-p}&0\\
0&0&0&0&0&0&0&1\\

\end{pmatrix}.
\end{equation}

Finally, for a collision between $B$ and $C$ we have that
\begin{equation}\label{w21}
\ket{g_{A},g_{B},g_{C}} \longrightarrow \ket{g_{A},g_{B},g_{C}},
\end{equation}
\begin{equation}\label{w22}
\ket{g_{A},g_{B},e_{C}} \longrightarrow\sqrt{1-p} \ket{g_{A},g_{B},e_{C}} -\sqrt{p} \ket{g_{A},e_{B},g_{C}},
\end{equation}
\begin{equation}\label{w23}
\ket{g_{A},e_{B},g_{C}} \longrightarrow \sqrt{1-p} \ket{g_{A},e_{B},g_{C}} +\sqrt{p} \ket{g_{A},g_{B},e_{C}},
\end{equation}
\begin{equation}\label{w24}
\ket{g_{A},e_{B},e_{C}} \longrightarrow \ket{g_{A},e_{B},e_{C}} ,
\end{equation}
\begin{equation}\label{w25}
\ket{e_{A},g_{B},g_{C}} \longrightarrow \ket{e_{A},g_{B},g_{C}} ,
\end{equation}
\begin{equation}\label{w26}
\ket{e_{A},g_{B},e_{C}} \longrightarrow \sqrt{1-p} \ket{e_{A},g_{B},e_{C}} -\sqrt{p} \ket{e_{A},e_{B},g_{C}},
\end{equation}
\begin{equation}\label{w27}
\ket{e_{A},e_{B},g_{C}} \longrightarrow\sqrt{1-p} \ket{e_{A},e_{B},g_{C}} + \sqrt{p}\ket{e_{A},g_{B},e_{C}},
\end{equation}
\begin{equation}\label{w28}
\ket{e_{A},e_{B},e_{C}} \longrightarrow \ket{e_{A},e_{B},e_{C}}.
\end{equation}
The unitary transformation matrix obtained from the above relations is

\begin{equation}\label{canal4}
\hat{U}_{BC}=\begin{pmatrix}
1&0&0&0&0&0&0&0\\ 0&\sqrt{1-p}&\sqrt{p}&0&0&0&0&0\\
0&-\sqrt{1-p}&\sqrt{p}&0&0&0&0&0\\
0&0&0&1&0&0&0&0\\
0&0&0&0&1&0&0&0\\
0&0&0&0&0&\sqrt{1-p}&\sqrt{p}&0\\
0&0&0&0&0&-\sqrt{p}&\sqrt{1-p}&0\\
0&0&0&0&0&0&0&1\\

\end{pmatrix}.
\end{equation}

\begin{figure}[h!]
	\includegraphics[scale=0.26]{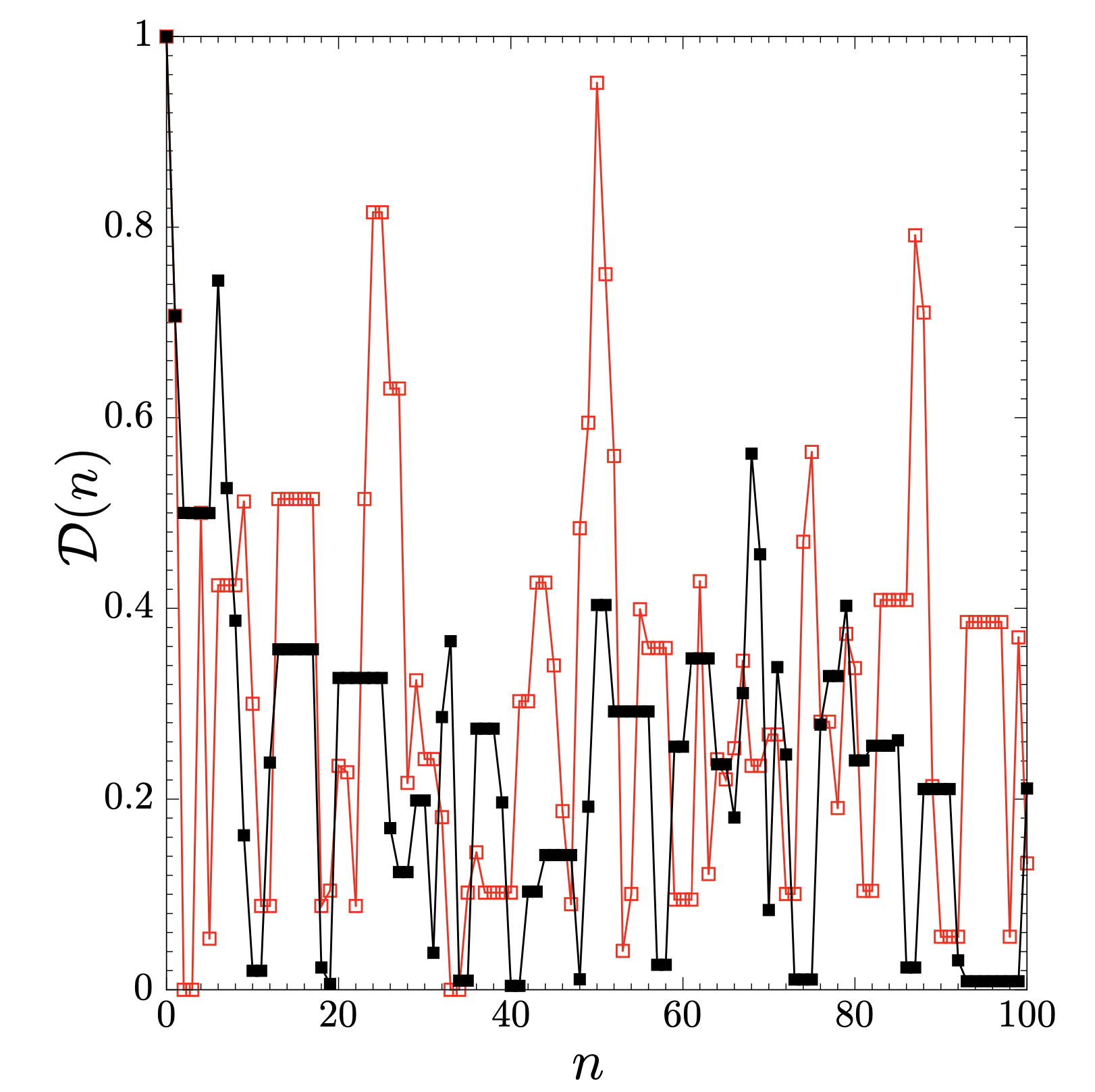}
	\centering
	\caption{Trace distance of the initially orthogonal states of $A$ as a function of the number of collisions for the two- (red open boxes) and three-qubit (solid black boxes) environment cases. For the initial state of all environment constituents, we used Eq.~(\ref{idensmatB}) with $w_{g}=0.8$ and $w_{e}=0.2$. We also used $p=0.5$ in both cases. These results indicate suppression of the non-Markovian behavior of the open quantum dynamics with increasing environment size.}
	\label{fig7}
\end{figure}

With these results, we can calculate the state evolution of our system $A$ for an arbitrary sequence of collisions involving $A$, $B$ and $C$. For instance, for a sequence of four collisions, in which the first occurs between $A$ and $C$, the second between $B$ and $C$, the third between $B$ and $C$, and the fourth between $A$ and $B$, we have that the state of $A$ becomes

\begin{equation}\label{3pevolution}
\hat{\rho}_{A}(4) = Tr_{BC}[\hat{U}_{AB}\hat{U}^2_{BC}\hat{U}_{AC}\hat{\rho}_{ABC}(0)\hat{U}^{\dagger}_{AC}\hat{U}^{\dagger 2}_{BC}\hat{U}^{\dagger}_{AB}],
\end{equation}
where $\hat{\rho}_{ABC}(0) = \hat{\rho}_{A}(0) \otimes \hat{\rho}_{B}(0) \otimes \hat{\rho}_{C}(0)$, and $Tr_{BC}$ denotes trace over the states of $B$ and $C$. This procedure can be extended for an arbitrary number of collisions, such that the properties of $A$ can be calculated upon evolution of the composite system. In order to study the presence of non-Markovianity in this case, in Fig.~\ref{fig7} we show the behavior of the dynamics of the trace distance between the orthogonal states of $A$ shown in Eqs.~(\ref{rho1}) and~(\ref{rho2}) for a generated random sequence of 100 collisions. We used the parameters $w_{g} = 0.8$, $w_{e} = 0.2$, and $p=0.5$. 

As can also be seen, in Fig.~\ref{fig7} the trace distance exhibits an oscillatory behavior, which indicates non-Markovianity, but on average it clearly assumes values smaller than those of the single-qubit environment case, Fig.~\ref{fig5}(a). This fact indicates reduction of the non-Markovian dynamical properties due to the addition of the extra qubit $C$ to the environment. For comparison, we also show in Fig.~\ref{fig7} the results of an extended calculation for the case in which the environment has three qubits, i.e., when a new thermal qubit $D$ (initially identical to $B$ and $C$) is added to the composite system. This result clearly shows that the trace distance still oscillates, but on average assuming values smaller than the two-qubit environment case. Details of the calculations of the three-qubit environment are provided in the Supplemental Material. 

In general, we see that with the dynamical interactions presented here, the addition of one ancilla to the environment is sufficient to cause a noticeable reduction in the backflow of information from the environment to the system. This decrease in the non-Markovian properties with increasing environment size is explained by the fact that larger environments contain more constituents to spread and store the absorbed information. In both the two- and three-qubit environment cases, we observe a chaotic behavior of the trace distance.

\begin{figure}[h!]
	\includegraphics[scale=0.26]{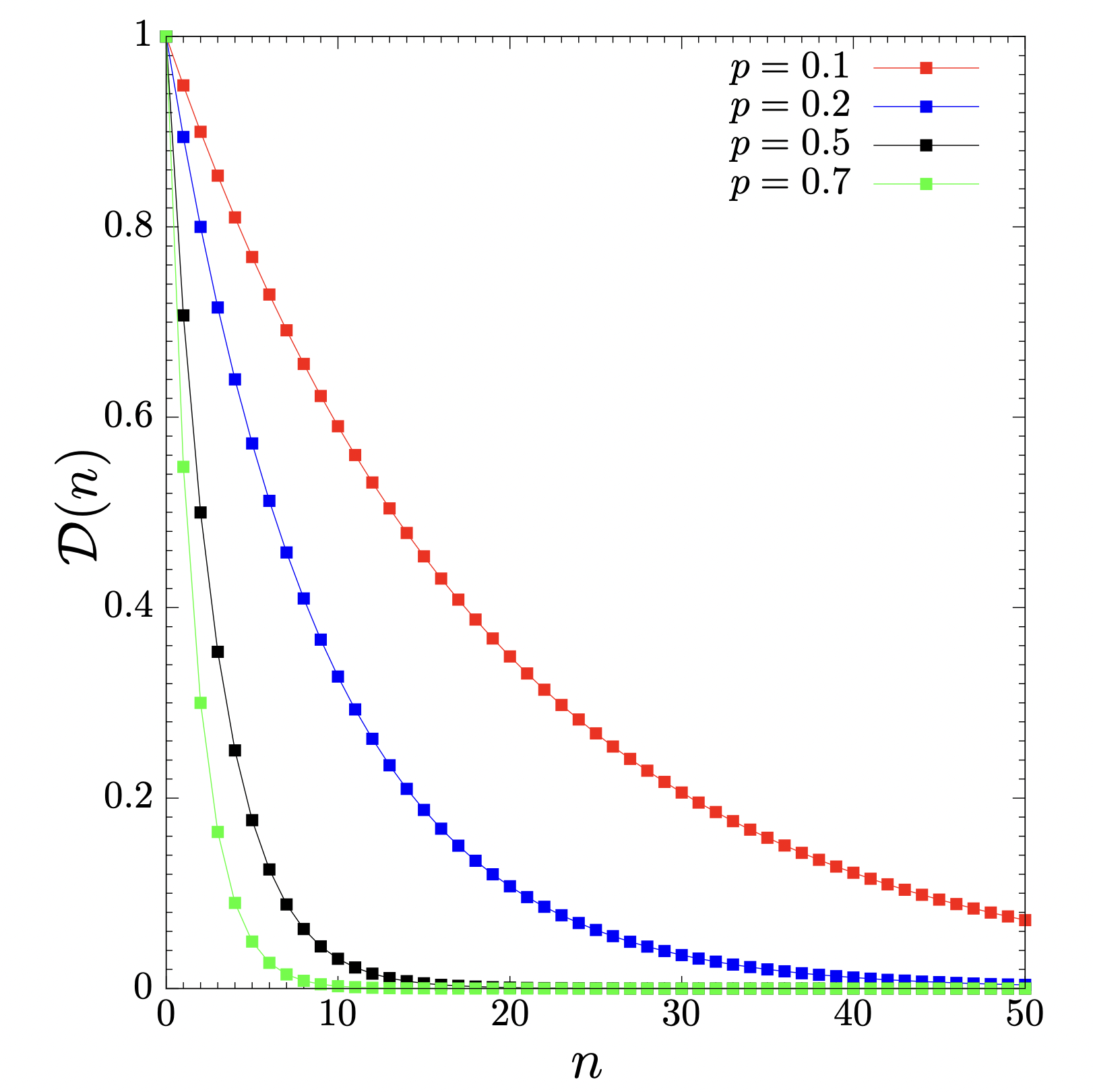}
	\centering
	\caption{Trace distance of the initially orthogonal states of $A$ as a function of the number of collisions when the environment comprises a large number of identical thermal qubits (ancillas). For the initial state of the ancillas, we used Eq.~(\ref{idensmatB}) with $w_{g}=0.8$ and $w_{e}=0.2$. We display $\mathcal{D} (n)$ for the cases in which the interaction probabilities are $p=0.1$, $p=0.2$, $p=0.5$ and $p=0.7$, where the monotonically decreasing behavior is a characteristic of Markovian quantum evolutions. Numerical calculations also showed that the system always thermalizes in this regime.}
	\label{fig8}
\end{figure}

As a last point, it is worth briefly commenting on the limit when the environment is made up of an infinitely large number of identical ancillas. In this case, we assume that $i$) the states of the system and the ancillas are initially uncorrelated, $ii$) the system collides only once with each ancilla, and $iii$) the ancillas that interact with the system do not interact with each other.  As such, the state of $A$ after $n$ collisions yields \cite{ciccarello}  
\begin{equation}\label{infpevolution}
\hat{\rho}_{A}(n)= \mathcal{E}\left[\cdot \cdot \cdot \mathcal{E}\left[\mathcal{E}[\hat{\rho}_{A}(0)]\right]\right] = \mathcal{E}^n\left[ \hat{\rho}_{A}(0) \right],
\end{equation}
where we define the map
\begin{equation}\label{infpevolution3}
\mathcal{E}[\hat{\rho}] = Tr_{env} [ \hat{U} (\hat{\rho} \otimes \hat{\rho}_{env}) \hat{U}^{\dagger} ],
\end{equation}
with $\hat{\rho}_{env} = w_{g}\ket{g}\bra{g} + w_{e}\ket{e}\bra{e}$ being the initial state of the ancillas, and $Tr_{env}$ the trace over the ancillary basis states, $\ket{g}$ and $\ket{e}$. The hypotheses $i$) to $iii$) underpin the main properties of a Markovian dynamics. In order to illustrate this effect, in Fig.~\ref{fig8} we show the evolution of the trace distance $\mathcal{D}(n)$ of the system states given in Eqs.~(\ref{rho1}) and~(\ref{rho2}) as a function of $n$, where we considered $w_{g} = 0.8$ and $w_{e} = 0.2$. We can see that $\mathcal{D}(n)$ is a monotonically decreasing function, so that Ineq.~(\ref{td}) is never fulfilled, and the bigger $p$ the faster the flow of information to the environment \cite{breuer3}. Our numerical calculations also showed that the system always thermalizes in this scenario, i.e., $\lim_{n \to \infty} \hat{\rho}_{A}(n) = \hat{\rho}_{env}$. This point is in accordance with Refs.~\cite{jevtic,mancino}.

\section{Conclusion}

In this work, we have studied memory effects in the dynamics of open quantum systems with collisional models, in which both the system of interest and the environment ancillas are considered as qubits. The dynamics of the system and the ancillas are dictated by a sequence of short pairwise unitary interactions, described by Eq.~(\ref{canal1}), which models the evolution of a two-level system in contact with a heat bath at non-zero temperature. In the simplest case of a single ancilla, we observed that the dynamics of the coherence of the system and the system-ancilla entanglement manifest either chaotic or regular oscillatory behavior, depending on the interaction probability $p$ of each collision. Orbit diagrams of the coherence dynamics showed the long-term behavior for a continuum of values of $p$, which revealed an interesting composition of regularity and chaos. 

We also investigated the system-environment information flow in order to learn whether, how, and to what extent quantum non-Markovian effects take place in the system evolution. To this end, we calculated the dynamics of the trace distance between two orthogonal initial states of the system, according to the criterion proposed in Ref. \cite{breuer3}. Our results showed the presence of very strong non-Markovianity in the single ancilla case, in which there is a permanent back and forth flow of information between system and ancilla. The dynamics of the trace distance also becomes chaotic or regular as the value of $p$ varies, and indicates that the initial information is allowed to return completely to the system after few collisions.

We then examined the changes in the non-Markovian behavior when the size of the environment increases, whereas both system-ancilla and ancilla-ancilla collisions are considered. When two ancillary qubits are present, the trace distance still oscillates, but a complete backflow of information is no longer observed. We also showed that this suppression in the backflow is even more prominent when there are three ancillary qubits. It is expected that this loss of memory effects continues as we keep adding ancillas to the environment up to the point when the trace distance becomes a monotonically decreasing function of the number of collisions, which means that the dynamics becomes Markovian. We believe the present results provide important insights on how the interaction parameters and environment size dictate a transition between Markovian and non-Markovian quantum behavior.

\section*{Acknowledgements}
        
The authors acknowledge support from Coordena{\c c}{\~a}o de Aperfei{\c c}oamento
de Pessoal de N{\'i}vel Superior (CAPES, Finance Code 001) and Conselho Nacional de Desenvolvimento Cient{\'i}fico e Tecnol{\'o}gico (CNPq). BLB acknowledges support from (CNPq, Grant No. 303451/2019-0), and PROPESQ/PRPG/UFPB (Project code PIA13177-2020).

\newpage

\section*{SUPPLEMENTAL MATERIAL}

In this Supplemental Material, we detail the procedure to calculate the dynamics of the state of the system in contact with three thermal ancillas, in which system-ancilla and ancilla-ancilla collisions are considered. The result allows the investigation of the non-Markovian properties presented in the main article.

\section*{Three-qubit environment}

This is the case in which four qubit particles are involved: the system $A$ and three environment constituents (ancillas) $B$, $C$ and $D$. As discussed in the main article, all particles have ground and excited states with energies $E_{g}$ and $E_{e}$, respectively. The initial state of $A$ is given by $\ket{\psi(0)} = a\ket{g_{A}}+b\ket{e_{A}}$, and the ancillary initial states are $\hat{\rho}_{B}(0) = w_{g}\ket{g_{B}}\bra{g_{B}} + w_{e}\ket{e_{B}}\bra{e_{B}}$, $\hat{\rho}_{C}(0) = w_{g}\ket{g_{C}}\bra{g_{C}} + w_{e}\ket{e_{C}}\bra{e_{C}}$ and $\hat{\rho}_{D}(0) = w_{g}\ket{g_{D}}\bra{g_{D}} + w_{e}\ket{e_{D}}\bra{e_{D}}$. The probabilistic weights obey the relation $w_{e} = w_{g} e^{-\beta(E_{e} - E_{g})}$, with $\beta$ the inverse temperature and $w_{g}+w_{e} = 1$. We also assume that the four particles start out in a separable state, $\hat{\rho}_{ABCD}(0) = \hat{\rho}_{A}(0) \otimes \hat{\rho}_{B}(0) \otimes \hat{\rho}_{C}(0)\otimes \hat{\rho}_{D}(0)$.

To account for quantum correlations involving the four particles, which are formed upon evolution, arbitrary qubit-qubit collisions must be described in the energy Hilbert space of $A$, $B$, $C$ and $D$ \cite{cohen}. We can span this 16-dimensional space with the following set of energy states:

\begin{equation}
\begin{aligned}
    M = \{\ket{g_{A},g_{B},g_{C},g_{D}} ; \ket{g_{A},g_{B},g_{C},e_{D}} ; \ket{g_{A},g_{B},e_{C},g_{D}} ; \ket{g_{A},g_{B},e_{C},e_{D}} ;\\
    \ket{g_{A},e_{B},g_{C},g_{D}} ; \ket{g_{A},e_{B},e_{C},e_{D}} ; \ket{g_{A},e_{B},e_{C},g_{D}} ; \ket{g_{A},e_{B},e_{C},e_{D}} ;\\
    \ket{e_{A},g_{B},g_{C},g_{D}} ;
    \ket{e_{A},g_{B},g_{C},e_{D}} ;
    \ket{e_{A},g_{B},e_{C},e_{D}} ;
    \ket{e_{A},g_{B},e_{C},e_{D}} ;\\
    \ket{e_{A},e_{B},g_{C},g_{D}} ;
    \ket{e_{A},e_{B},g_{C},e_{D}} ; 
    \ket{e_{A},e_{B},e_{C},g_{D}} ;
    \ket{e_{A},e_{B},e_{C},e_{D}}  \}.
\end{aligned}
\end{equation}
For convenience of notation, we alternatively relabel the above basis states, respectively, as 

\begin{equation}
\begin{aligned}
    M = \{\ket{1} ; \ket{2} ; \ket{3} ; \ket{4} ;
    \ket{5} ; \ket{5} ; \ket{7} ; \ket{8} ;
    \ket{9} ; \ket{10} ; \ket{11} ; \ket{12} ;
    \ket{13} ; \ket{14} ; \ket{15} ; \ket{16} \},
\end{aligned}
\end{equation}
which will also be used. We now proceed to investigate the quantum maps that describe all possible qubit-qubit interactions \cite{nielsen2,preskill2,khatri2}.

\subsection{System-ancilla collisions}

By following the same interaction rules depicted in the main article, if we are interested for example in the evolution of the four particles when $A$ collides with $B$, we must extend the relations (9) to (12) to the form

\begin{equation}
\begin{aligned}
     \ket{g_{A},g_{B},g_{C},g_{D}} \rightarrow \ket{g_{A},g_{B},g_{C},g_{D}},
\end{aligned}
\end{equation}

\begin{equation}
\begin{aligned}
     \ket{g_{A},g_{B},e_{C},g_{D}} \rightarrow \ket{g_{A},g_{B},e_{C},g_{D}},
\end{aligned}
\end{equation}

\begin{equation}
\begin{aligned}
    \ket{g_{A},g_{B},g_{C},e_{D}} \rightarrow \ket{g_{A},g_{B},g_{C},e_{D}},
\end{aligned}
\end{equation}

\begin{equation}
\begin{aligned}
     \ket{g_{A},g_{B},e_{C},e_{D}} \rightarrow \ket{g_{A},g_{B},e_{C},e_{D}},
\end{aligned}
\end{equation}

\begin{equation}
\begin{aligned}
     \ket{e_{A},g_{B},g_{C},g_{D}} \rightarrow \sqrt{1-p}\ket{e_{A},g_{B},g_{C},g_{D}} + \sqrt{p}\ket{g_{A},e_{B},g_{C},g_{D}}, 
\end{aligned}
\end{equation}

\begin{equation}
\begin{aligned}
     \ket{e_{A},g_{B},e_{C},g_{D}} \rightarrow \sqrt{1-p}\ket{e_{A},g_{B},e_{C},g_{D}} + \sqrt{p}\ket{g_{A},e_{B},e_{C},g_{D}}, 
\end{aligned}
\end{equation}

\begin{equation}
\begin{aligned}
     \ket{e_{A},g_{B},g_{C},e_{D}} \rightarrow \sqrt{1-p}\ket{e_{A},g_{B},g_{C},e_{D}} + \sqrt{p}\ket{g_{A},e_{B},g_{C},e_{D}}, 
\end{aligned}
\end{equation}

\begin{equation}
\begin{aligned}
     \ket{e_{A},g_{B},e_{C},e_{D}} \rightarrow \sqrt{1-p}\ket{e_{A},g_{B},e_{C},e_{D}} + \sqrt{p}\ket{g_{A},e_{B},e_{C},e_{D}}, 
\end{aligned}
\end{equation}

\begin{equation}
\begin{aligned}
    \ket{g_{A},e_{B},g_{C},g_{D}} \rightarrow \sqrt{1-p}\ket{g_{A},e_{B},g_{C},g_{D}} - \sqrt{p}\ket{e_{A},g_{B},g_{C},g_{D}}, 
\end{aligned}
\end{equation}

\begin{equation}
\begin{aligned}
    \ket{g_{A},e_{B},e_{C},g_{D}} \rightarrow \sqrt{1-p}\ket{g_{A},e_{B},e_{C},g_{D}} - \sqrt{p}\ket{e_{A},g_{B},e_{C},g_{D}}, 
\end{aligned}
\end{equation}

\begin{equation}
\begin{aligned}
    \ket{g_{A},e_{B},g_{C},e_{D}} \rightarrow \sqrt{1-p}\ket{g_{A},e_{B},g_{C},e_{D}} - \sqrt{p}\ket{e_{A},g_{B},g_{C},e_{D}},
\end{aligned}
\end{equation}

\begin{equation}
\begin{aligned}
    \ket{g_{A},e_{B},e_{C},e_{D}} \rightarrow \sqrt{1-p}\ket{g_{A},e_{B},e_{C},e_{D}} - \sqrt{p}\ket{e_{A},g_{B},e_{C},e_{D}}, 
\end{aligned}
\end{equation}

\begin{equation}
\begin{aligned}
    \ket{e_{A},e_{B},g_{C},g_{D}} \rightarrow \ket{e_{A},e_{B},g_{C},g_{D}},
\end{aligned}
\end{equation}

\begin{equation}
\begin{aligned}
    \ket{e_{A},e_{B},e_{C},g_{D}} \rightarrow \ket{e_{A},e_{B},e_{C},g_{D}},
\end{aligned}
\end{equation}

\begin{equation}
\begin{aligned}
    \ket{e_{A},e_{B},g_{C},e_{D}} \rightarrow \ket{e_{A},e_{B},g_{C},e_{D}},
\end{aligned}
\end{equation}

\begin{equation}
\begin{aligned}
    \ket{e_{A},e_{B},e_{C},e_{D}} \rightarrow \ket{e_{A},e_{B},e_{C},e_{D}}.
\end{aligned}
\end{equation}
Note that the energy states of particles $C$ and $D$ have no influence on this dynamics. With the above relations, we can write the unitary that describes a collision between $A$ and $B$:

\begin{equation}
\begin{split}
\hat{U}_{AB} = &\ket{1}\bra{1} + \ket{2}\bra{2} + \ket{3}\bra{3} + \ket{4}\bra{4} +\sqrt{1-p}\ket{5}\bra{5}\\
& - \sqrt{p}\ket{9}\bra{5} + \sqrt{1-p}\ket{6}\bra{6} - \sqrt{p}\ket{10}\bra{6}\\ 
&+\sqrt{1-p}\ket{7}\bra{7} - \sqrt{p}\ket{11}\bra{7}+ \sqrt{1-p}\ket{8}\bra{8} \\
&- \sqrt{p} \ket{12}\bra{8} 
+\sqrt{1-p}\ket{9}\bra{9} + \sqrt{p}\ket{5}\bra{9}  \\
&+ \sqrt{1-p}\ket{10}\bra{10}+ \sqrt{p} \ket{6}\bra{10} 
+\sqrt{1-p}\ket{11}\bra{11} \\
&+ \sqrt{p}\ket{7}\bra{11} + \sqrt{1-p}\ket{12}\bra{12} + \sqrt{p} \ket{8}\bra{12} \\
&+\ket{13}\bra{13} + \ket{14}\bra{14} + \ket{15}\bra{15} + \ket{16}\bra{16},
\end{split}
\end{equation}
which provides the unitary matrix

\begin{equation}
\setcounter{MaxMatrixCols}{16}
    \hat{U}_{AB} = 
    \begin{bmatrix}   
    1 & 0 & 0 & 0 & 0 & 0 & 0 & 0 & 0 & 0 & 0 & 0 & 0 & 0 & 0 & 0 \\
    0 & 1 & 0 & 0 & 0 & 0 & 0 & 0 & 0 & 0 & 0 & 0 & 0 & 0 & 0 & 0 \\
    0 & 0 & 1 & 0 & 0 & 0 & 0 & 0 & 0 & 0 & 0 & 0 & 0 & 0 & 0 & 0 \\
    0 & 0 & 0 & 1 & 0 & 0 & 0 & 0 & 0 & 0 & 0 & 0 & 0 & 0 & 0 & 0 \\
    0 & 0 & 0 & 0 & \sqrt{1-p} & 0 & 0 & 0 & \sqrt{p} & 0 & 0 & 0 & 0 & 0 & 0 & 0 \\
    0 & 0 & 0 & 0 & 0 & \sqrt{1-p} & 0 & 0 & 0 & \sqrt{p} & 0 & 0 & 0 & 0 & 0 & 0 \\
    0 & 0 & 0 & 0 & 0 & 0 & \sqrt{1-p} & 0 & 0 & 0 & \sqrt{p} & 0 & 0 & 0 & 0 & 0 \\
    0 & 0 & 0 & 0 & 0 & 0 & 0 & \sqrt{1-p} & 0 & 0 & 0 & \sqrt{p} & 0 & 0 & 0 & 0 \\
    0 & 0 & 0 & 0 & -\sqrt{p} & 0 & 0 & 0 & \sqrt{1-p} & 0 & 0 & 0 & 0 & 0 & 0 & 0 \\
    0 & 0 & 0 & 0 & 0 & -\sqrt{p} & 0 & 0 & 0 & \sqrt{1-p} & 0 & 0 & 0 & 0 & 0 & 0 \\
    0 & 0 & 0 & 0 & 0 & 0 & -\sqrt{p} & 0 & 0 & 0 & \sqrt{1-p} & 0 & 0 & 0 & 0 & 0 \\
    0 & 0 & 0 & 0 & 0 & 0 & 0 & -\sqrt{p} & 0 & 0 & 0 & \sqrt{1-p} & 0 & 0 & 0 & 0 \\
    0 & 0 & 0 & 0 & 0 & 0 & 0 & 0 & 0 & 0 & 0 & 0 & 1 & 0 & 0 & 0 \\
    0 & 0 & 0 & 0 & 0 & 0 & 0 & 0 & 0 & 0 & 0 & 0 & 0 & 1 & 0 & 0 \\
    0 & 0 & 0 & 0 & 0 & 0 & 0 & 0 & 0 & 0 & 0 & 0 & 0 & 0 & 1 & 0 \\
    0 & 0 & 0 & 0 & 0 & 0 & 0 & 0 & 0 & 0 & 0 & 0 & 0 & 0 & 0 & 1 \\
    \end{bmatrix}.  
\end{equation}

Similarly, for a collision involving A and C, we have that

\begin{equation}
\begin{aligned}
     \ket{g_{A},g_{B},g_{C},g_{D}} \rightarrow \ket{g_{A},g_{B},g_{C},g_{D}},
\end{aligned}
\end{equation}

\begin{equation}
\begin{aligned}
     \ket{g_{A},e_{B},g_{C},g_{D}} \rightarrow \ket{g_{A},e_{B},g_{C},g_{D}},
\end{aligned}
\end{equation}

\begin{equation}
\begin{aligned}
     \ket{g_{A},g_{B},g_{C},e_{D}} \rightarrow \ket{g_{A},g_{B},g_{C},e_{D}},
\end{aligned}
\end{equation}

\begin{equation}
\begin{aligned}
     \ket{g_{A},e_{B},g_{C},e_{D}} \rightarrow \ket{g_{A},e_{B},g_{C},e_{D}},
\end{aligned}
\end{equation}

\begin{equation}
\begin{aligned}
     \ket{e_{A},g_{B},g_{C},g_{D}} \rightarrow \sqrt{1-p}\ket{e_{A},g_{B},g_{C},g_{D}} + \sqrt{p}\ket{g_{A},g_{B},e_{C},g_{D}},
\end{aligned}
\end{equation}

\begin{equation}
\begin{aligned}
     \ket{e_{A},e_{B},g_{C},g_{D}} \rightarrow \sqrt{1-p}\ket{e_{A},e_{B},g_{C},g_{D}} + \sqrt{p}\ket{g_{A},e_{B},e_{C},g_{D}},
\end{aligned}
\end{equation}

\begin{equation}
\begin{aligned}
     \ket{e_{A},g_{B},g_{C},e_{D}} \rightarrow \sqrt{1-p}\ket{e_{A},g_{B},g_{C},e_{D}} + \sqrt{p}\ket{g_{A},g_{B},e_{C},e_{D}},
\end{aligned}
\end{equation}

\begin{equation}
\begin{aligned}
     \ket{e_{A},e_{B},g_{C},e_{D}} \rightarrow \sqrt{1-p}\ket{e_{A},e_{B},g_{C},e_{D}} + \sqrt{p}\ket{g_{A},e_{B},e_{C},e_{D}},
\end{aligned}
\end{equation}

\begin{equation}
\begin{aligned}
     \ket{g_{A},g_{B},e_{C},g_{D}} \rightarrow \sqrt{1-p}\ket{g_{A},g_{B},e_{C},g_{D}} - \sqrt{p}\ket{e_{A},g_{B},g_{C},g_{D}},
\end{aligned}
\end{equation}

\begin{equation}
\begin{aligned}
    \ket{g_{A},e_{B},e_{C},g_{D}} \rightarrow \sqrt{1-p}\ket{g_{A},e_{B},e_{C},g_{D}} - \sqrt{p}\ket{e_{A},e_{B},g_{C},g_{D}},
\end{aligned}
\end{equation}

\begin{equation}
\begin{aligned}
    \ket{g_{A},g_{B},e_{C},e_{D}} \rightarrow \sqrt{1-p}\ket{g_{A},g_{B},e_{C},e_{D}} - \sqrt{p}\ket{e_{A},g_{B},g_{C},e_{D}},
\end{aligned}
\end{equation}

\begin{equation}
\begin{aligned}
    \ket{g_{A},e_{B},e_{C},e_{D}} \rightarrow \sqrt{1-p}\ket{g_{A},e_{B},e_{C},e_{D}} - \sqrt{p}\ket{e_{A},e_{B},g_{C},e_{D}},
\end{aligned}
\end{equation}

\begin{equation}
\begin{aligned}
    \ket{e_{A},g_{B},e_{C},g_{D}} \rightarrow \ket{e_{A},g_{B},e_{C},g_{D}},
\end{aligned}
\end{equation}

\begin{equation}
\begin{aligned}
    \ket{e_{A},e_{B},e_{C},g_{D}} \rightarrow \ket{e_{A},e_{B},e_{C},g_{D}},
\end{aligned}
\end{equation}

\begin{equation}
\begin{aligned}
    \ket{e_{A},g_{B},e_{C},e_{D}} \rightarrow \ket{e_{A},g_{B},e_{C},e_{D}},
\end{aligned}
\end{equation}

\begin{equation}
\begin{aligned}
    \ket{e_{A},e_{B},e_{C},e_{D}} \rightarrow \ket{e_{A},e_{B},e_{C},e_{D}},
\end{aligned}
\end{equation}

whose equivalent unitary is given by

\begin{equation}
\begin{split}
\hat{U}_{AC} = &\ket{1}\bra{1} + \ket{2}\bra{2} + \sqrt{1-p}\ket{3}\bra{3} -\sqrt{p}\ket{9}\bra{3} +\sqrt{1-p}\ket{4}\bra{4} \\
& - \sqrt{p} \ket{10}\bra{4} + \ket{5}\bra{5} + \ket{6}\bra{6} +\sqrt{1-p}\ket{7}\bra{7}- \sqrt{p}\ket{13}\bra{7} \\
& + \sqrt{1-p}\ket{8}\bra{8} - \sqrt{p} \ket{14}\bra{8}+\sqrt{1-p}\ket{9}\bra{9} + \sqrt{p}\ket{3}\bra{9} \\
& + \sqrt{1-p}\ket{10}\bra{10} + \sqrt{p} \ket{4}\bra{10} +\ket{11}\bra{11} + \ket{12}\bra{12} \\
&+ \sqrt{1-p}\ket{13}\bra{13} + \sqrt{p} \ket{7}\bra{13} +\sqrt{1-p}\ket{14}\bra{14}\\
& + \sqrt{p}\ket{8}\bra{14} + \ket{15}\bra{15} + \ket{16}\bra{16}.
\end{split}
\end{equation}

This gives us the matrix

\begin{equation}
\setcounter{MaxMatrixCols}{16}
    \hat{U}_{AC} = 
    \begin{bmatrix}   
    1 & 0 & 0 & 0 & 0 & 0 & 0 & 0 & 0 & 0 & 0 & 0 & 0 & 0 & 0 & 0 \\
    0 & 1 & 0 & 0 & 0 & 0 & 0 & 0 & 0 & 0 & 0 & 0 & 0 & 0 & 0 & 0 \\
    0 & 0 & \sqrt{1-p} & 0 & 0 & 0 & 0 & 0 & \sqrt{p} & 0 & 0 & 0 & 0 & 0 & 0 & 0 \\
    0 & 0 & 0 & \sqrt{1-p} & 0 & 0 & 0 & 0 & 0 & \sqrt{p} & 0 & 0 & 0 & 0 & 0 & 0 \\
    0 & 0 & 0 & 0 & 1 & 0 & 0 & 0 & 0 & 0 & 0 & 0 & 0 & 0 & 0 & 0 \\
    0 & 0 & 0 & 0 & 0 & 1 & 0 & 0 & 0 & 0 & 0 & 0 & 0 & 0 & 0 & 0 \\
    0 & 0 & 0 & 0 & 0 & 0 & \sqrt{1-p} & 0 & 0 & 0 & 0 & 0 & \sqrt{p} & 0 & 0 & 0 \\
    0 & 0 & 0 & 0 & 0 & 0 & 0 & \sqrt{1-p} & 0 & 0 & 0 & 0 & 0 & \sqrt{p} & 0 & 0 \\
    0 & 0 & -\sqrt{p} & 0 & 0 & 0 & 0 & 0 & \sqrt{1-p} & 0 & 0 & 0 & 0 & 0 & 0 & 0 \\
    0 & 0 & 0 & -\sqrt{p} & 0 & 0 & 0 & 0 & 0 & \sqrt{1-p} & 0 & 0 & 0 & 0 & 0 & 0 \\
    0 & 0 & 0 & 0 & 0 & 0 & 0 & 0 & 0 & 0 & 1 & 0 & 0 & 0 & 0 & 0 \\
    0 & 0 & 0 & 0 & 0 & 0 & 0 & 0 & 0 & 0 & 0 & 1 & 0 & 0 & 0 & 0 \\
    0 & 0 & 0 & 0 & 0 & 0 & -\sqrt{p} & 0 & 0 & 0 & 0 & 0 & \sqrt{1-p} & 0 & 0 & 0 \\
    0 & 0 & 0 & 0 & 0 & 0 & 0 & -\sqrt{p} & 0 & 0 & 0 & 0 & 0 & \sqrt{1-p} & 0 & 0 \\
    0 & 0 & 0 & 0 & 0 & 0 & 0 & 0 & 0 & 0 & 0 & 0 & 0 & 0 & 1 & 0 \\
    0 & 0 & 0 & 0 & 0 & 0 & 0 & 0 & 0 & 0 & 0 & 0 & 0 & 0 & 0 & 1 \\
    \end{bmatrix}.  
\end{equation}

Finally, for a collision involving $A$ and $D$, we have that

\begin{equation}
\begin{aligned}
     \ket{g_{A},g_{B},g_{C},g_{D}} \rightarrow \ket{g_{A},g_{B},g_{C},g_{D}},
\end{aligned}
\end{equation}

\begin{equation}
\begin{aligned}
     \ket{g_{A},e_{B},g_{C},g_{D}} \rightarrow \ket{g_{A},e_{B},g_{C},g_{D}}, 
\end{aligned}
\end{equation}

\begin{equation}
\begin{aligned}
     \ket{g_{A},g_{B},e_{C},g_{D}} \rightarrow \ket{g_{A},g_{B},e_{C},g_{D}},
\end{aligned}
\end{equation}

\begin{equation}
\begin{aligned}
     \ket{g_{A},e_{B},e_{C},g_{D}} \rightarrow \ket{g_{A},e_{B},e_{C},g_{D}}, 
\end{aligned}
\end{equation}

\begin{equation}
\begin{aligned}
     \ket{e_{A},g_{B},g_{C},g_{D}} \rightarrow \sqrt{1-p}\ket{e_{A},g_{B},g_{C},g_{D}} + \sqrt{p}\ket{g_{A},g_{B},g_{C},e_{D}}, 
\end{aligned}
\end{equation}

\begin{equation}
\begin{aligned}
     \ket{e_{A},e_{B},g_{C},g_{D}} \rightarrow \sqrt{1-p}\ket{e_{A},e_{B},g_{C},g_{D}} + \sqrt{p}\ket{g_{A},e_{B},g_{C},e_{D}}, 
\end{aligned}
\end{equation}

\begin{equation}
\begin{aligned}
     \ket{e_{A},g_{B},e_{C},g_{D}} \rightarrow \sqrt{1-p}\ket{e_{A},g_{B},e_{C},g_{D}} + \sqrt{p}\ket{g_{A},g_{B},e_{C},e_{D}}, 
\end{aligned}
\end{equation}

\begin{equation}
\begin{aligned}
     \ket{e_{A},e_{B},e_{C},g_{D}} \rightarrow \sqrt{1-p}\ket{e_{A},e_{B},e_{C},g_{D}} + \sqrt{p}\ket{g_{A},e_{B},e_{C},e_{D}}, 
\end{aligned}
\end{equation}

\begin{equation}
\begin{aligned}
     \ket{g_{A},g_{B},g_{C},e_{D}} \rightarrow \sqrt{1-p}\ket{g_{A},g_{B},g_{C},e_{D}} - \sqrt{p}\ket{e_{A},g_{B},g_{C},g_{D}}, 
\end{aligned}
\end{equation}

\begin{equation}
\begin{aligned}
     \ket{g_{A},e_{B},g_{C},e_{D}} \rightarrow \sqrt{1-p}\ket{g_{A},e_{B},g_{C},e_{D}} - \sqrt{p}\ket{e_{A},e_{B},g_{C},g_{D}},
\end{aligned}
\end{equation}

\begin{equation}
\begin{aligned}
    \ket{g_{A},g_{B},e_{C},e_{D}} \rightarrow \sqrt{1-p}\ket{g_{A},g_{B},e_{C},e_{D}} - \sqrt{p}\ket{e_{A},g_{B},e_{C},g_{D}},
\end{aligned}
\end{equation}
    
\begin{equation}
\begin{aligned}
    \ket{g_{A},e_{B},e_{C},e_{D}} \rightarrow \sqrt{1-p}\ket{g_{A},e_{B},e_{C},e_{D}} - \sqrt{p}\ket{e_{A},e_{B},e_{C},g_{D}},
\end{aligned}
\end{equation}

\begin{equation}
\begin{aligned}
    \ket{e_{A},g_{B},g_{C},e_{D}} \rightarrow \ket{e_{A},g_{B},g_{C},e_{D}}, 
\end{aligned}
\end{equation}

\begin{equation}
\begin{aligned}
    \ket{e_{A},e_{B},g_{C},e_{D}} \rightarrow \ket{e_{A},e_{B},g_{C},e_{D}},
\end{aligned}
\end{equation}

\begin{equation}
\begin{aligned}
    \ket{e_{A},g_{B},e_{C},e_{D}} \rightarrow \ket{e_{A},g_{B},e_{C},e_{D}},
\end{aligned}
\end{equation}

\begin{equation}
\begin{aligned}
    \ket{e_{A},e_{B},e_{C},e_{D}} \rightarrow \ket{e_{A},e_{B},e_{C},e_{D}}.
\end{aligned}
\end{equation}
The following unitary is then obtained

\begin{equation}
\begin{split}
\hat{U}_{AD} = &\ket{1}\bra{1} + \sqrt{1-p}\ket{2}\bra{2} - \sqrt{p} \ket{9}\bra{2} + \ket{3}\bra{3} 
+\sqrt{1-p}\ket{4}\bra{4}\\
&- \sqrt{p} \ket{11}\bra{4} + \ket{5}\bra{5} + \sqrt{1-p}\ket{6}\bra{6} 
-\sqrt{p}\ket{13}\bra{6}+ \ket{7}\bra{7} \\
&+ \sqrt{1-p}\ket{8}\bra{8} - \sqrt{p} \ket{15}\bra{8} +\sqrt{1-p}\ket{9}\bra{9} + \sqrt{p}\ket{2}\bra{9}\\
& + \ket{10}\bra{10} + \sqrt{1-p} \ket{11}\bra{11} +\sqrt{p}\ket{4}\bra{11} + \ket{12}\bra{12} \\
&+ \sqrt{1-p}\ket{13}\bra{13} + \sqrt{p} \ket{6}\bra{13} +\ket{14}\bra{14} + \sqrt{1-p}\ket{15}\bra{15}\\
& + \sqrt{p}\ket{8}\bra{15} + \ket{16}\bra{16},
\end{split}
\end{equation}
which yields

\begin{equation}
\setcounter{MaxMatrixCols}{16}
    \hat{U}_{AD} = 
    \begin{bmatrix}   
    1 & 0 & 0 & 0 & 0 & 0 & 0 & 0 & 0 & 0 & 0 & 0 & 0 & 0 & 0 & 0 \\
    0 & \sqrt{1-p} & 0 & 0 & 0 & 0 & 0 & 0 & \sqrt{p} & 0 & 0 & 0 & 0 & 0 & 0 & 0 \\
    0 & 0 & 1 & 0 & 0 & 0 & 0 & 0 & 0 & 0 & 0 & 0 & 0 & 0 & 0 & 0 \\
    0 & 0 & 0 & \sqrt{1-p} & 0 & 0 & 0 & 0 & 0 & 0 & \sqrt{p} & 0 & 0 & 0 & 0 & 0 \\
    0 & 0 & 0 & 0 & 1 & 0 & 0 & 0 & 0 & 0 & 0 & 0 & 0 & 0 & 0 & 0 \\
    0 & 0 & 0 & 0 & 0 & \sqrt{1-p} & 0 & 0 & 0 & 0 & 0 & 0 & \sqrt{p} & 0 & 0 & 0 \\
    0 & 0 & 0 & 0 & 0 & 0 & 1 & 0 & 0 & 0 & 0 & 0 & 0 & 0 & 0 & 0 \\
    0 & 0 & 0 & 0 & 0 & 0 & 0 & \sqrt{1-p} & 0 & 0 & 0 & 0 & 0 & 0 & \sqrt{p} & 0 \\
    0 & -\sqrt{p} & 0 & 0 & 0 & 0 & 0 & 0 & \sqrt{1-p} & 0 & 0 & 0 & 0 & 0 & 0 & 0 \\
    0 & 0 & 0 & 0 & 0 & 0 & 0 & 0 & 0 & 1 & 0 & 0 & 0 & 0 & 0 & 0 \\
    0 & 0 & 0 & -\sqrt{p} & 0 & 0 & 0 & 0 & 0 & 0 & \sqrt{1-p} & 0 & 0 & 0 & 0 & 0 \\
    0 & 0 & 0 & 0 & 0 & 0 & 0 & 0 & 0 & 0 & 0 & 1 & 0 & 0 & 0 & 0 \\
    0 & 0 & 0 & 0 & 0 & -\sqrt{p} & 0 & 0 & 0 & 0 & 0 & 0 & \sqrt{1-p} & 0 & 0 & 0 \\
    0 & 0 & 0 & 0 & 0 & 0 & 0 & 0 & 0 & 0 & 0 & 0 & 0 & 1 & 0 & 0 \\
    0 & 0 & 0 & 0 & 0 & 0 & 0 & -\sqrt{p} & 0 & 0 & 0 & 0 & 0 & 0 & \sqrt{1-p} & 0 \\
    0 & 0 & 0 & 0 & 0 & 0 & 0 & 0 & 0 & 0 & 0 & 0 & 0 & 0 & 0 & 1 \\
    \end{bmatrix}.  
\end{equation}

\subsection{Ancilla-ancilla collisions}

We now proceed to investigate the dynamics that result from ancilla-ancilla collisions. We begin with the collision between $B$ and $C$, whose evolution is described by the following relations: 

\begin{equation}
\begin{aligned}
     \ket{g_{A},g_{B},g_{C},g_{D}} \rightarrow \ket{g_{A},g_{B},g_{C},g_{D}},
\end{aligned}
\end{equation}

\begin{equation}
\begin{aligned}
     \ket{e_{A},g_{B},g_{C},g_{D}} \rightarrow \ket{e_{A},g_{B},g_{C},g_{D}},
\end{aligned}
\end{equation}

\begin{equation}
\begin{aligned}
     \ket{g_{A},g_{B},g_{C},e_{D}} \rightarrow \ket{g_{A},g_{B},g_{C},e_{D}},
\end{aligned}
\end{equation}

\begin{equation}
\begin{aligned}
     \ket{e_{A},g_{B},g_{C},e_{D}} \rightarrow \ket{e_{A},g_{B},g_{C},e_{D}}, 
\end{aligned}
\end{equation}

\begin{equation}
\begin{aligned}
     \ket{g_{A},e_{B},g_{C},g_{D}} \rightarrow \sqrt{1-p}\ket{g_{A},e_{B},g_{C},g_{D}} + \sqrt{p}\ket{g_{A},g_{B},e_{C},g_{D}}, 
\end{aligned}
\end{equation}

\begin{equation}
\begin{aligned}
     \ket{e_{A},e_{B},g_{C},g_{D}} \rightarrow \sqrt{1-p}\ket{e_{A},e_{B},g_{C},g_{D}} + \sqrt{p}\ket{e_{A},g_{B},e_{C},g_{D}}, 
\end{aligned}
\end{equation}

\begin{equation}
\begin{aligned}
     \ket{g_{A},e_{B},g_{C},e_{D}} \rightarrow \sqrt{1-p}\ket{g_{A},e_{B},g_{C},e_{D}} + \sqrt{p}\ket{g_{A},g_{B},e_{C},e_{D}}, 
\end{aligned}
\end{equation}

\begin{equation}
\begin{aligned}
     \ket{e_{A},e_{B},g_{C},e_{D}} \rightarrow \sqrt{1-p}\ket{e_{A},e_{B},g_{C},e_{D}} + \sqrt{p}\ket{e_{A},g_{B},e_{C},e_{D}}, 
\end{aligned}
\end{equation}

\begin{equation}
\begin{aligned}
     \ket{g_{A},g_{B},e_{C},g_{D}} \rightarrow \sqrt{1-p}\ket{g_{A},g_{B},e_{C},g_{D}} - \sqrt{p}\ket{g_{A},e_{B},g_{C},g_{D}}, 
\end{aligned}
\end{equation}

\begin{equation}
\begin{aligned}
     \ket{e_{A},g_{B},e_{C},g_{D}} \rightarrow \sqrt{1-p}\ket{e_{A},g_{B},e_{C},g_{D}} - \sqrt{p}\ket{e_{A},e_{B},g_{C},g_{D}}, 
\end{aligned}
\end{equation}

\begin{equation}
\begin{aligned}
     \ket{g_{A},g_{B},e_{C},e_{D}} \rightarrow \sqrt{1-p}\ket{g_{A},g_{B},e_{C},e_{D}} - \sqrt{p}\ket{g_{A},e_{B},g_{C},e_{D}}, 
\end{aligned}
\end{equation}

\begin{equation}
\begin{aligned}
    \ket{e_{A},g_{B},e_{C},e_{D}} \rightarrow \sqrt{1-p}\ket{e_{A},g_{B},e_{C},e_{D}} - \sqrt{p}\ket{e_{A},e_{B},g_{C},e_{D}}, 
\end{aligned}
\end{equation}

\begin{equation}
\begin{aligned}
    \ket{g_{A},e_{B},e_{C},g_{D}} \rightarrow \ket{g_{A},e_{B},e_{C},g_{D}}, 
\end{aligned}
\end{equation}

\begin{equation}
\begin{aligned}
    \ket{e_{A},e_{B},e_{C},g_{D}} \rightarrow \ket{e_{A},e_{B},e_{C},g_{D}}, 
\end{aligned}
\end{equation}

\begin{equation}
\begin{aligned}
    \ket{g_{A},e_{B},e_{C},e_{D}} \rightarrow \ket{g_{A},e_{B},e_{C},e_{D}}, 
\end{aligned}
\end{equation}

\begin{equation}
\begin{aligned}
    \ket{e_{A},e_{B},e_{C},e_{D}} \rightarrow \ket{e_{A},e_{B},e_{C},e_{D}}.
\end{aligned}
\end{equation}
They allow us to write the unitary operator

\begin{equation}
\begin{split}
\hat{U}_{BC} = &\ket{1}\bra{1} + \ket{2}\bra{2} + \sqrt{1-p} \ket{3}\bra{3} - \sqrt{p} \ket{5}\bra{3}+\sqrt{1-p}\ket{4}\bra{4} \\
& - \sqrt{p} \ket{6}\bra{4} + \sqrt{1-p}\ket{5}\bra{5} + \sqrt{p}\ket{3}\bra{5}+\sqrt{1-p}\ket{6}\bra{6}  \\
&+ \sqrt{p} \ket{4}\bra{6} + \ket{7}\bra{7} + \ket{8}\bra{8}+\ket{9}\bra{9} + \ket{10}\bra{10}+ \sqrt{1-p}\ket{11}\bra{11}  \\
& - \sqrt{p} \ket{13}\bra{11} +\sqrt{1-p}\ket{12}\bra{12} - \sqrt{p}\ket{14}\bra{12} + \sqrt{1-p}\ket{13}\bra{13}\\
& + \sqrt{p} \ket{11}\bra{13}+\sqrt{1-p}\ket{14}\bra{14} + \sqrt{p}\ket{12}\bra{14} + \ket{15}\bra{15} + \ket{16}\bra{16},
\end{split}
\end{equation}
which provides the matrix

\begin{equation}
\setcounter{MaxMatrixCols}{16}
    \hat{U}_{BC} = 
    \begin{bmatrix}   
    1 & 0 & 0 & 0 & 0 & 0 & 0 & 0 & 0 & 0 & 0 & 0 & 0 & 0 & 0 & 0 \\
    0 & 1 & 0 & 0 & 0 & 0 & 0 & 0 & 0 & 0 & 0 & 0 & 0 & 0 & 0 & 0 \\
    0 & 0 & \sqrt{1-p} & 0 & \sqrt{p} & 0 & 0 & 0 & 0 & 0 & 0 & 0 & 0 & 0 & 0 & 0 \\
    0 & 0 & 0 & \sqrt{1-p} & 0 & \sqrt{p} & 0 & 0 & 0 & 0 & 0 & 0 & 0 & 0 & 0 & 0 \\
    0 & 0 & -\sqrt{p} & 0 & \sqrt{1-p} & 0 & 0 & 0 & 0 & 0 & 0 & 0 & 0 & 0 & 0 & 0 \\
    0 & 0 & 0 & -\sqrt{p} & 0 & \sqrt{1-p} & 0 & 0 & 0 & 0 & 0 & 0 & 0 & 0 & 0 & 0 \\
    0 & 0 & 0 & 0 & 0 & 0 & 1 & 0 & 0 & 0 & 0 & 0 & 0 & 0 & 0 & 0 \\
    0 & 0 & 0 & 0 & 0 & 0 & 0 & 1 & 0 & 0 & 0 & 0 & 0 & 0 & 0 & 0 \\
    0 & 0 & 0 & 0 & 0 & 0 & 0 & 0 & 1 & 0 & 0 & 0 & 0 & 0 & 0 & 0 \\
    0 & 0 & 0 & 0 & 0 & 0 & 0 & 0 & 0 & 1 & 0 & 0 & 0 & 0 & 0 & 0 \\
    0 & 0 & 0 & 0 & 0 & 0 & 0 & 0 & 0 & 0 & \sqrt{1-p} & 0 & \sqrt{p} & 0 & 0 & 0 \\
    0 & 0 & 0 & 0 & 0 & 0 & 0 & 0 & 0 & 0 & 0 & \sqrt{1-p} & 0 & \sqrt{p} & 0 & 0 \\
    0 & 0 & 0 & 0 & 0 & 0 & 0 & 0 & 0 & 0 & -\sqrt{p} & 0 & \sqrt{1-p} & 0 & 0 & 0 \\
    0 & 0 & 0 & 0 & 0 & 0 & 0 & 0 & 0 & 0 & 0 & -\sqrt{p} & 0 & \sqrt{1-p} & 0 & 0 \\
    0 & 0 & 0 & 0 & 0 & 0 & 0 & 0 & 0 & 0 & 0 & 0 & 0 & 0 & 1 & 0 \\
    0 & 0 & 0 & 0 & 0 & 0 & 0 & 0 & 0 & 0 & 0 & 0 & 0 & 0 & 0 & 1 \\
    \end{bmatrix}.  
\end{equation}

In a similar fashion, for a collision involving $B$ and $D$, we have that

\begin{equation}
\begin{aligned}
     \ket{g_{A},g_{B},g_{C},g_{D}} \rightarrow \ket{g_{A},g_{B},g_{C},g_{D}},
\end{aligned}
\end{equation}

\begin{equation}
\begin{aligned}
     \ket{e_{A},g_{B},g_{C},g_{D}} \rightarrow \ket{e_{A},g_{B},g_{C},g_{D}}, 
\end{aligned}
\end{equation}

\begin{equation}
\begin{aligned}
     \ket{g_{A},g_{B},e_{C},g_{D}} \rightarrow \ket{g_{A},g_{B},e_{C},g_{D}},
\end{aligned}
\end{equation}

\begin{equation}
\begin{aligned}
     \ket{e_{A},g_{B},e_{C},g_{D}} \rightarrow \ket{e_{A},g_{B},e_{C},g_{D}}, 
\end{aligned}
\end{equation}

\begin{equation}
\begin{aligned}
     \ket{g_{A},e_{B},g_{C},g_{D}} \rightarrow \sqrt{1-p}\ket{g_{A},e_{B},g_{C},g_{D}} + \sqrt{p}\ket{g_{A},g_{B},g_{C},e_{D}}, 
\end{aligned}
\end{equation}

\begin{equation}
\begin{aligned}
     \ket{e_{A},e_{B},g_{C},g_{D}} \rightarrow \sqrt{1-p}\ket{e_{A},e_{B},g_{C},g_{D}} + \sqrt{p}\ket{e_{A},g_{B},g_{C},e_{D}}, 
\end{aligned}
\end{equation}

\begin{equation}
\begin{aligned}
     \ket{g_{A},e_{B},e_{C},g_{D}} \rightarrow \sqrt{1-p}\ket{g_{A},e_{B},e_{C},g_{D}} + \sqrt{p}\ket{g_{A},g_{B},e_{C},e_{D}}, 
\end{aligned}
\end{equation}

\begin{equation}
\begin{aligned}
     \ket{e_{A},e_{B},e_{C},g_{D}} \rightarrow \sqrt{1-p}\ket{e_{A},e_{B},e_{C},g_{D}} + \sqrt{p}\ket{e_{A},g_{B},e_{C},e_{D}}, 
\end{aligned}
\end{equation}

\begin{equation}
\begin{aligned}
     \ket{g_{A},g_{B},g_{C},e_{D}} \rightarrow \sqrt{1-p}\ket{g_{A},g_{B},g_{C},e_{D}} - \sqrt{p}\ket{g_{A},e_{B},g_{C},g_{D}}, 
\end{aligned}
\end{equation}

\begin{equation}
\begin{aligned}
     \ket{e_{A},g_{B},g_{C},e_{D}} \rightarrow \sqrt{1-p}\ket{e_{A},g_{B},g_{C},e_{D}} - \sqrt{p}\ket{e_{A},e_{B},g_{C},g_{D}}, 
\end{aligned}
\end{equation}

\begin{equation}
\begin{aligned}
     \ket{g_{A},g_{B},e_{C},e_{D}} \rightarrow \sqrt{1-p}\ket{g_{A},g_{B},e_{C},e_{D}} - \sqrt{p}\ket{g_{A},e_{B},e_{C},g_{D}}, 
\end{aligned}
\end{equation}

\begin{equation}
\begin{aligned}
     \ket{e_{A},g_{B},e_{C},e_{D}} \rightarrow \sqrt{1-p}\ket{e_{A},g_{B},e_{C},e_{D}} - \sqrt{p}\ket{e_{A},e_{B},e_{C},g_{D}}, 
\end{aligned}
\end{equation}

\begin{equation}
\begin{aligned}
    \ket{g_{A},e_{B},g_{C},e_{D}} \rightarrow \ket{g_{A},e_{B},g_{C},e_{D}}, 
\end{aligned}
\end{equation}

\begin{equation}
\begin{aligned}
    \ket{e_{A},e_{B},g_{C},e_{D}} \rightarrow \ket{e_{A},e_{B},g_{C},e_{D}}, 
\end{aligned}
\end{equation}

\begin{equation}
\begin{aligned}
    \ket{g_{A},e_{B},e_{C},e_{D}} \rightarrow \ket{g_{A},e_{B},e_{C},e_{D}}, 
\end{aligned}
\end{equation}

\begin{equation}
\begin{aligned}
    \ket{e_{A},e_{B},e_{C},e_{D}} \rightarrow \ket{e_{A},e_{B},e_{C},e_{D}}.
\end{aligned}
\end{equation}
The associated operator is then

\begin{equation}
\begin{split}
\hat{U}_{BD} = &\ket{1}\bra{1} + \sqrt{1-p}\ket{2}\bra{2} - \sqrt{p} \ket{5}\bra{2} + \ket{3}\bra{3} +\sqrt{1-p}\ket{4}\bra{4} \\
&- \sqrt{p} \ket{7}\bra{4} + \sqrt{1-p}\ket{5}\bra{5} + \sqrt{p}\ket{2}\bra{5} +\ket{6}\bra{6} + \sqrt{1-p} \ket{7}\bra{7}\\
 &+ \sqrt{p} \ket{4}\bra{7} + \ket{8}\bra{8}+\ket{9}\bra{9} + \sqrt{1-p}\ket{10}\bra{10} - \sqrt{p}\ket{13}\bra{10} \\
 &+ \ket{11}\bra{11}+\sqrt{1-p}\ket{12}\bra{12} - \sqrt{p}\ket{15}\bra{12}+ \sqrt{1-p}\ket{13}\bra{13}  \\
 &+ \sqrt{p} \ket{10}\bra{13}
+\ket{14}\bra{14} + \sqrt{1-p}\ket{15}\bra{15} + \sqrt{p}\ket{12}\bra{15} + \ket{16}\bra{16},
\end{split}
\end{equation}
with the corresponding matrix

\begin{equation}
\setcounter{MaxMatrixCols}{16}
    \hat{U}_{BD} = 
    \begin{bmatrix}   
    1 & 0 & 0 & 0 & 0 & 0 & 0 & 0 & 0 & 0 & 0 & 0 & 0 & 0 & 0 & 0 \\
    0 & \sqrt{1-p} & 0 & 0 & \sqrt{p} & 0 & 0 & 0 & 0 & 0 & 0 & 0 & 0 & 0 & 0 & 0 \\
    0 & 0 & 1 & 0 & 0 & 0 & 0 & 0 & 0 & 0 & 0 & 0 & 0 & 0 & 0 & 0 \\
    0 & 0 & 0 & \sqrt{1-p} & 0 & 0 & \sqrt{p} & 0 & 0 & 0 & 0 & 0 & 0 & 0 & 0 & 0 \\
    0 & -\sqrt{p} & 0 & 0 & \sqrt{1-p} & 0 & 0 & 0 & 0 & 0 & 0 & 0 & 0 & 0 & 0 & 0 \\
    0 & 0 & 0 & 0 & 0 & 1 & 0 & 0 & 0 & 0 & 0 & 0 & 0 & 0 & 0 & 0 \\
    0 & 0 & 0 & -\sqrt{p} & 0 & 0 & \sqrt{1-p} & 0 & 0 & 0 & 0 & 0 & 0 & 0 & 0 & 0 \\
    0 & 0 & 0 & 0 & 0 & 0 & 0 & 1 & 0 & 0 & 0 & 0 & 0 & 0 & 0 & 0 \\
    0 & 0 & 0 & 0 & 0 & 0 & 0 & 0 & 1 & 0 & 0 & 0 & 0 & 0 & 0 & 0 \\
    0 & 0 & 0 & 0 & 0 & 0 & 0 & 0 & 0 & \sqrt{1-p} & 0 & 0 & \sqrt{p} & 0 & 0 & 0 \\
    0 & 0 & 0 & 0 & 0 & 0 & 0 & 0 & 0 & 0 & 1 & 0 & 0 & 0 & 0 & 0 \\
    0 & 0 & 0 & 0 & 0 & 0 & 0 & 0 & 0 & 0 & 0 & \sqrt{1-p} & 0 & 0 & \sqrt{p} & 0 \\
    0 & 0 & 0 & 0 & 0 & 0 & 0 & 0 & 0 & -\sqrt{p} & 0 & 0 & \sqrt{1-p} & 0 & 0 & 0 \\
    0 & 0 & 0 & 0 & 0 & 0 & 0 & 0 & 0 & 0 & 0 & 0 & 0 & 1 & 0 & 0 \\
    0 & 0 & 0 & 0 & 0 & 0 & 0 & 0 & 0 & 0 & 0 & -\sqrt{p} & 0 & 0 & \sqrt{1-p} & 0 \\
    0 & 0 & 0 & 0 & 0 & 0 & 0 & 0 & 0 & 0 & 0 & 0 & 0 & 0 & 0 & 1 \\
    \end{bmatrix}.  
\end{equation}

Lastly, we have that a collision involving $C$ and $D$ provides the relations 

\begin{equation}
\begin{aligned}
     \ket{g_{A},g_{B},g_{C},g_{D}} \rightarrow \ket{g_{A},g_{B},g_{C},g_{D}},
\end{aligned}
\end{equation}

\begin{equation}
\begin{aligned}
     \ket{e_{A},g_{B},g_{C},g_{D}} \rightarrow \ket{e_{A},g_{B},g_{C},g_{D}}, 
\end{aligned}
\end{equation}

\begin{equation}
\begin{aligned}
     \ket{g_{A},e_{B},g_{C},g_{D}} \rightarrow \ket{g_{A},e_{B},g_{C},g_{D}},
\end{aligned}
\end{equation}

\begin{equation}
\begin{aligned}
     \ket{e_{A},e_{B},g_{C},g_{D}} \rightarrow \ket{e_{A},e_{B},g_{C},g_{D}}, 
\end{aligned}
\end{equation}

\begin{equation}
\begin{aligned}
     \ket{g_{A},g_{B},e_{C},g_{D}} \rightarrow \sqrt{1-p}\ket{g_{A},g_{B},e_{C},g_{D}} + \sqrt{p}\ket{g_{A},g_{B},g_{C},e_{D}}, 
\end{aligned}
\end{equation}

\begin{equation}
\begin{aligned}
     \ket{e_{A},g_{B},e_{C},g_{D}} \rightarrow \sqrt{1-p}\ket{e_{A},g_{B},e_{C},g_{D}} + \sqrt{p}\ket{e_{A},g_{B},g_{C},e_{D}}, 
\end{aligned}
\end{equation}

\begin{equation}
\begin{aligned}
     \ket{g_{A},e_{B},e_{C},g_{D}} \rightarrow \sqrt{1-p}\ket{g_{A},e_{B},e_{C},g_{D}} + \sqrt{p}\ket{g_{A},e_{B},g_{C},e_{D}}, 
\end{aligned}
\end{equation}

\begin{equation}
\begin{aligned}
     \ket{e_{A},e_{B},e_{C},g_{D}} \rightarrow \sqrt{1-p}\ket{e_{A},e_{B},e_{C},g_{D}} + \sqrt{p}\ket{e_{A},e_{B},g_{C},e_{D}}, 
\end{aligned}
\end{equation}

\begin{equation}
\begin{aligned}
     \ket{g_{A},g_{B},g_{C},e_{D}} \rightarrow \sqrt{1-p}\ket{g_{A},g_{B},g_{C},e_{D}} - \sqrt{p}\ket{g_{A},g_{B},e_{C},g_{D}}, 
\end{aligned}
\end{equation}

\begin{equation}
\begin{aligned}
     \ket{e_{A},g_{B},g_{C},e_{D}} \rightarrow \sqrt{1-p}\ket{e_{A},g_{B},g_{C},e_{D}} - \sqrt{p}\ket{e_{A},g_{B},e_{C},g_{D}}, 
\end{aligned}
\end{equation}

\begin{equation}
\begin{aligned}
     \ket{g_{A},e_{B},g_{C},e_{D}} \rightarrow \sqrt{1-p}\ket{g_{A},e_{B},g_{C},e_{D}} - \sqrt{p}\ket{g_{A},e_{B},e_{C},g_{D}}, 
\end{aligned}
\end{equation}

\begin{equation}
\begin{aligned}
     \ket{e_{A},e_{B},g_{C},e_{D}} \rightarrow \sqrt{1-p}\ket{e_{A},e_{B},g_{C},e_{D}} - \sqrt{p}\ket{e_{A},e_{B},e_{C},g_{D}}, 
\end{aligned}
\end{equation}

\begin{equation}
\begin{aligned}
     \ket{g_{A},g_{B},e_{C},e_{D}} \rightarrow \ket{g_{A},g_{B},e_{C},e_{D}}, 
\end{aligned}
\end{equation}

\begin{equation}
\begin{aligned}
    \ket{e_{A},g_{B},e_{C},e_{D}} \rightarrow \ket{e_{A},g_{B},e_{C},e_{D}}, 
\end{aligned}
\end{equation}
    
\begin{equation}
\begin{aligned}
    \ket{g_{A},e_{B},e_{C},e_{D}} \rightarrow \ket{g_{A},e_{B},e_{C},e_{D}}, 
\end{aligned}
\end{equation}

\begin{equation}
\begin{aligned}
    \ket{e_{A},e_{B},e_{C},e_{D}} \rightarrow \ket{e_{A},e_{B},e_{C},e_{D}}, 
\end{aligned}
\end{equation}
from which we write the unitary operator

\begin{equation}
\begin{split}
\hat{U}_{CD} = &\ket{1}\bra{1} + \sqrt{1-p}\ket{2}\bra{2} - \sqrt{p} \ket{3}\bra{2} + \sqrt{1-p}\ket{3}\bra{3} +\sqrt{p}\ket{2}\bra{3} + \ket{4}\bra{4}\\
& + \ket{5}\bra{5} + \sqrt{1-p}\ket{6}\bra{6} -\sqrt{p}\ket{7}\bra{6} + \sqrt{1-p} \ket{7}\bra{7}+ \sqrt{p} \ket{6}\bra{7}  \\
&+ \ket{8}\bra{8}+\ket{9}\bra{9} + \sqrt{1-p}\ket{10}\bra{10} - \sqrt{p}\ket{11}\bra{10} + \sqrt{1-p}\ket{11}\bra{11} \\
&+\sqrt{p}\ket{10}\bra{11} + \ket{12}\bra{12} + \ket{13}\bra{13} + \sqrt{1-p} \ket{14}\bra{14} -\sqrt{p}\ket{15}\bra{14}\\
& + \sqrt{1-p}\ket{15}\bra{15} + \sqrt{p}\ket{14}\bra{15} + \ket{16}\bra{16},
\end{split}
\end{equation}
and the corresponding unitary matrix

\begin{equation}
\setcounter{MaxMatrixCols}{16}
    \hat{U}_{CD} = 
    \begin{bmatrix}   
    1 & 0 & 0 & 0 & 0 & 0 & 0 & 0 & 0 & 0 & 0 & 0 & 0 & 0 & 0 & 0 \\
    0 & \sqrt{1-p} & \sqrt{p} & 0 & 0 & 0 & 0 & 0 & 0 & 0 & 0 & 0 & 0 & 0 & 0 & 0 \\
    0 & -\sqrt{p} & \sqrt{1-p} & 0 & 0 & 0 & 0 & 0 & 0 & 0 & 0 & 0 & 0 & 0 & 0 & 0 \\
    0 & 0 & 0 & 1 & 0 & 0 & 0 & 0 & 0 & 0 & 0 & 0 & 0 & 0 & 0 & 0 \\
    0 & 0 & 0 & 0 & 1 & 0 & 0 & 0 & 0 & 0 & 0 & 0 & 0 & 0 & 0 & 0 \\
    0 & 0 & 0 & 0 & 0 & \sqrt{1-p} & \sqrt{p} & 0 & 0 & 0 & 0 & 0 & 0 & 0 & 0 & 0 \\
    0 & 0 & 0 & 0 & 0 & -\sqrt{p} & \sqrt{1-p} & 0 & 0 & 0 & 0 & 0 & 0 & 0 & 0 & 0 \\
    0 & 0 & 0 & 0 & 0 & 0 & 0 & 1 & 0 & 0 & 0 & 0 & 0 & 0 & 0 & 0 \\
    0 & 0 & 0 & 0 & 0 & 0 & 0 & 0 & 1 & 0 & 0 & 0 & 0 & 0 & 0 & 0 \\
    0 & 0 & 0 & 0 & 0 & 0 & 0 & 0 & 0 & \sqrt{1-p} & \sqrt{p} & 0 & 0 & 0 & 0 & 0 \\
    0 & 0 & 0 & 0 & 0 & 0 & 0 & 0 & 0 & -\sqrt{p} & \sqrt{1-p} & 0 & 0 & 0 & 0 & 0 \\
    0 & 0 & 0 & 0 & 0 & 0 & 0 & 0 & 0 & 0 & 0 & 1 & 0 & 0 & 0 & 0 \\
    0 & 0 & 0 & 0 & 0 & 0 & 0 & 0 & 0 & 0 & 0 & 0 & 1 & 0 & 0 & 0 \\
    0 & 0 & 0 & 0 & 0 & 0 & 0 & 0 & 0 & 0 & 0 & 0 & 0 & \sqrt{1-p} & \sqrt{p} & 0 \\
    0 & 0 & 0 & 0 & 0 & 0 & 0 & 0 & 0 & 0 & 0 & 0 & 0 & -\sqrt{p} & \sqrt{1-p} & 0 \\
    0 & 0 & 0 & 0 & 0 & 0 & 0 & 0 & 0 & 0 & 0 & 0 & 0 & 0 & 0 & 1 \\
    \end{bmatrix}.  
\end{equation}

\section*{Example}

With the results above, we can calculate the state evolution of our system $A$ for an arbitrary sequence of collisions involving the four particles. For example, in a sequence of seven collisions involving the pairs of particles: $A$-$C$, $B$-$C$, $A$-$D$, $B$-$D$, $A$-$B$, $A$-$B$, and $C$-$D$, we have that the state of $A$ becomes
\begin{equation}\label{4pevolution}
\hat{\rho}_{A}(7) = Tr_{BCD}[\hat{U}_{CD}\hat{U}^{2}_{AB}\hat{U}_{BD}\hat{U}_{AD} \hat{U}_{BC}\hat{U}_{AC}
\hat{\rho}_{ABCD}(0)\hat{U}^{\dagger}_{AC}\hat{U}^{\dagger}_{BC}\hat{U}^{\dagger}_{AD}\hat{U}^{\dagger}_{BD}\hat{U}^{\dagger 2}_{AB}\hat{U}^{\dagger}_{CD}],
\end{equation}
where $Tr_{BCD}$ denotes trace over the states of $B$, $C$ and $D$. This procedure can be extended for an arbitrary number of collisions, such that the properties of $A$ can be calculated from the evolution of the composite system. Fig.~7 of the main article illustrated the non-Markovian behavior of the dynamics of the system in contact with three thermal ancillas, as presented here. That result was calculated with basis on the state of $A$ obtained from a generated random sequence of 100 collisions.


\begin{thebibliography}{7}
\renewcommand{\baselinestretch}{2.0}

\bibitem{ladd} T. D. Ladd, F. Jelezko, R. Laflamme, Y. Nakamura, C. Monroe, and J. L. O'Brien, Nature (London) {\bf 464}, 45 (2010).

\bibitem{arute} F. Arute {\it et al.}, Nature (London) {\bf 574}, 505 (2019).

\bibitem{wu} Y. Wu {\it et al.}, Phys. Rev. Lett. {\bf 127}, 180501 (2021).

\bibitem{mishra} V. K. Mishra, {\it An Introduction to Quantum Communication} (Momentum Press, New York, NY, 2016).

\bibitem{simon} C. Simon, Towards a global quantum network, Nat. Photonics {\bf 11}, 678 (2017).

\bibitem{taylor} J. M. Taylor, P. Cappellaro, L. Childress, L. Jiang, D. Budker, P. R. Hemmer, A. Yacoby, R. Walsworth, and
M. D. Lukin, Nat. Phys. {\bf 4}, 810 (2008).

\bibitem{bera} M. L. Bera, M. Lewenstein, and M. N. Bera,
npj Quantum Inf. {\bf 7}, 31 (2021).

\bibitem{brand} K. Brandner, M. Bauer, and U. Seifert, Phys. Rev. Lett. {\bf 119}, 170602 (2017).

\bibitem{klaers} J. Klaers, S. Faelt, A. Imamoglu, and E. Togan, Phys. Rev. X {\bf 7}, 031044 (2017).

\bibitem{nielsen} M. A. Nielsen and I. L. Chuang, {\it Quantum Computation
and Quantum Information}, Cambridge Series on Information and the Natural Sciences (Cambridge University Press,
Cambridge, 2000).

\bibitem{breuer} H.-P. Breuer and F. Petruccione, {\it The Theory of Open Quantum Systems} (Oxford University Press, Oxford, 2007).

\bibitem{rivas} A. Rivas and S. F. Huelga, {\it Open Quantum Systems} (Springer Berlin Heidelberg, Berlin, Germany, 2012)

\bibitem{breuer2} H-P. Breuer, E.-M. Laine, J. Piilo, and B. Vacchini, Rev. Mod. Phys. {\bf 88}, 021002 (2016).

\bibitem{vega} I. de Vega and D. Alonso, Rev. Mod. Phys. {\bf 89}, 015001 (2017).

\bibitem{cli} C. Li, G. Guo, and J. Piilo, EPL {\bf 127}, 50001 (2019).

\bibitem{cli2} C. Li, G. Guo, and J. Piilo, EPL {\bf 128}, 30001 (2020).

\bibitem{byli} B. Bylicka, D. Chru\'{c}in\'{s}ki, and S. Maniscalco, Sci. Rep. {\bf 4}, 5720 (2014).

\bibitem{white} G. A. L. White, C. D. Hill, F. A. Pollock, L. C. L. Hollenberg, and K. Modi,
Nat. Commun. {\bf 11}, 6301 (2020).

\bibitem{liu} B.-H. Liu, L. Li, Y.-F. Huang, C.-F. Li, G.-C. Guo, E.-M. Laine, H.-P. Breuer, and J. Piilo,  Nat. Phys. {\bf 7} 931 (2011).

\bibitem{ciccarello} F. Ciccarello, S. Lorenzo, V. Giovannetti, and G. M. Palma, Phys. Rep. {\bf 954}, 1 (2022).

\bibitem{bernardes} N. K. Bernardes, A. R. R. Carvalho, C. H. Monken, and M. F. Santos, Phys. Rev. A {\bf 90}, 032111 (2014). 

\bibitem{khatri}  S. Khatri, K. Sharma, and M. M. Wilde, Phys. Rev. A {\bf 102}, 012401 (2020).

\bibitem{bose} S. Bose, Phys. Rev. Lett. {\bf 91}, 207901 (2003). 

\bibitem{goold} J. Goold, M. Paternostro, K. Modi, Phys. Rev. Lett. {\bf 114}, 060602 (2015).

\bibitem{chirolli} L. Chirolli and G. Burkard, Adv. Phys. {\bf 57}, 225 (2008).

\bibitem{zou} W.-J. Zou, Yu.-H. Li, S.-C. Wang, Y. Cao, J.-G. Ren, J. Yin,
C.-Z. Peng, X.-B. Wang, and J.-W. Pan, Phys. Rev. A {\bf 95},
042342 (2017).

\bibitem{bernardo} B. L. Bernardo
Phys. Rev. E {\bf 104}, 044111 (2021). 

\bibitem{maci} K. Macieszczak, M. Guta, I. Lesanovsky, and J. P. Garrahan,
Phys. Rev. Lett. {\bf 116}, 240404 (2016).


\bibitem{boite} A. Le Boit\'{e}, M.-J. Hwang, and M. B. Plenio, Phys. Rev. A {\bf 95},
023829 (2017).

\bibitem{valenti} D. Valenti, A. Carollo, and B. Spagnolo, Phys. Rev. A {\bf 97}, 042109
(2018).

\bibitem{bernardo2} B. L. Bernardo
Phys. Rev. E {\bf 102}, 062152 (2020). 

\bibitem{khan} K. Khan, J. S. Ara\'{u}jo, W. F. Magalh\~{a}es, G. H. Aguilar, and B. L. Bernardo, Quantum Sci. Technol. {\bf 7} 045010 (2022).

\bibitem{baum} T. Baumgratz, M. Cramer, and M. B. Plenio, Phys. Rev. Lett. {\bf 113}, 140401 (2014).

\bibitem{strogatz} S. H. Strogatz, {\it Nonlinear Dynamics and Chaos} (CRC Press, Boca Raton, FL, 2018)

\bibitem{vidal} G. Vidal and R. F. Werner, Phys. Rev. A {\bf 65}, 032314 (2002).

\bibitem{horodecki} R. Horodecki, P. Horodecki, M. Horodecki, and K. Horodecki, Rev. Mod. Phys. {\bf 81}, 865 (2009).


\bibitem{breuer3} H-P. Breuer, E.-M. Laine, J. Piilo, Phys. Rev. Lett. {\bf 103}, 210401 (2009).


\bibitem{jevtic} S. Jevtic, D. Newman, T. Rudolph, and T. M. Stace, Phys. Rev. A {\bf 91}, 012331 (2015).

\bibitem{mancino} L. Mancino, M. Sbroscia, I. Gianani, E. Roccia, and M. Barbieri,  Phys. Rev. Lett. {\bf 118}, 130502 (2017).





\end{thebibliography}

\begin{thebibliography}{11}

\bibitem {cohen} C. Cohen-Tannoudji, B. Diu and F. Laloe {\it Quantum Mechanics: Basic Concepts, Tools, and Applications; Angular Momentum, Spin, and Approximation; Fermions, Bosons, Photons, Correlations, and Entanglement}, Vol. 1 (WILEY-VCH Verlag GmbH \& Co. KGaA, Germany, 2020).
 
\bibitem {nielsen2} M. A. Nielsen and I. L. Chuang, {\it Quantum Computation
and Quantum Information}, Cambridge Series on Information and the Natural Sciences (Cambridge University Press,
Cambridge, 2000).
 
\bibitem{preskill2} J. Preskill, Lecture Notes for Physics 229: {\it Quantum Information
and Computation}, Vol. 16 (California Institute of Technology,
1998). 

\bibitem{khatri2}  S. Khatri, K. Sharma, and M. M. Wilde, Phys. Rev. A {\bf 102}, 012401 (2020).

\end{thebibliography}
\end{document}